\documentstyle [epsf]{mns}
\def\be{\begin{equation}}
\def\ee{\end{equation}}
\def\ba{\begin{eqnarray}}
\def\ea{\end{eqnarray}}

\begin{document}

\title[Structure of Dark Matter Halos]
{\bf The Structure and Dynamical Evolution of Dark Matter Halos}
\author[G.Tormen et al.]
{Giuseppe Tormen$^{1,2,3}$, Fran\c{c}ois R. Bouchet$^{1}$ and Simon
D. M. White$^{3}$ \\
$^1$Institut D'Astrophysique de Paris, 98 bis Boulevard Arago, 75014
Paris - FRANCE \\
$^2$Institute of Astronomy, University of Cambridge, Madingley Road, 
Cambridge CB3 0HA - UK\\
$^3$Max Planck Institute f\"{u}r Astrophysik, Karl-Schwarzschild-Strasse 1, 
85740 Garching bei M\"{u}nchen - GERMANY\\
\smallskip
Email: bepi@mpa-garching.mpg.de, bouchet@iap.fr, swhite@mpa-garching.mpg.de}

\pubyear{1997}

\maketitle

\begin{abstract}
We use $N$-body simulations to investigate the structure and 
dynamical evolution of dark matter halos in clusters of galaxies. 
Our sample consists of nine massive halos from an 
Einstein-De Sitter universe with scale free power spectrum 
and spectral index $n = -1$. Halos are resolved by 20000 
particles each, on average, and have a dynamical resolution 
of 20-25 kpc, as shown by extensive tests.
Large scale tidal fields are included up to a scale $L=150$ Mpc 
using background particles.
We find that the halo formation process can be characterized by the 
alternation of two dynamical configurations: a {\em merging} phase 
and a {\em relaxation} phase, defined by their signature on the 
evolution of the total mass and root mean square ({\em rms}) velocity.
Halos spend on average one third of their evolution in the merging
phase and two thirds in the relaxation phase.
Using this definition, we study the density profiles and show how they 
change during the halo dynamical history.
In particular, we find that the {\em average} density profiles of 
our halos are fitted by the Navarro, Frenk \& White (1995) analytical 
model with an {\em rms} residual of 17\% between the virial radius $R_v$ and
$0.01 R_v$. The Hernquist (1990) analytical density profiles fits the 
same halos with an {\em rms} residual of 26\%.
The trend with mass of the scale radius of these fits is marginally
consistent with that found by Cole \& Lacey (1996): compared to their 
results our halos are more centrally concentrated, and the relation
between scale radius and halo mass is slightly steeper. We find a 
moderately large scatter in this relation, due both to dynamical 
evolution within halos and to fluctuations in the halo population.
We analyze the dynamical equilibrium of our halos using the Jeans' 
equation, and find that on average they are approximately in 
equilibrium within their virial radius.
Finally, we find that the projected mass profiles of our simulated 
halos are in very good agreement with the profiles of three rich galaxy 
clusters derived from strong and weak gravitational lensing observations.

\end{abstract}


\section{Introduction}\label{sec:intro}

Observational studies of galaxy clusters are providing ever more
data that need theoretical interpretation in 
order to understand cluster formation and evolution. In current 
cosmological models the mass and dynamics of galaxy clusters are 
dominated by some kind of non-baryonic, {\em dark} matter, which 
interacts with ordinary baryonic matter only through gravity. 
In studies focussing on the dynamics of galaxy clusters, they can 
thus be regarded as halos made of collisionless dark matter. 
From the theoretical point of view one can study the structure of 
dark matter halos both analytically and numerically. Much of the 
analytical work done so far is based on the {\em secondary infall} 
paradigm, (Gunn \& Gott 1972). The simplest version of this picture 
considers an initial point mass, which acts as a nonlinear seed, 
surrounded by an homogeneous uniformly expanding universe.
Matter around the seed slows down due to its gravitational attraction, 
and eventually falls back in concentric spherical shells with purely 
radial motions. Calculations based on this model predict that the 
density profile of the virialized halo should scale as 
$\rho(r) \propto r^{-9/4}$.
Self similar solutions were found by Fillmore \& Goldreich (1984) 
and by Bertschinger (1985).
Hoffman \& Shaham (1985) applied an extension of this idea to the 
gravitational instability theory of hierarchical clustering. In their 
calculations they assumed a Gaussian random field of initial density 
perturbation with scale-free power spectra: $P(k) \propto k^n$. They 
found that the virialized structures originating from density peaks 
should have density profiles whose shape depends on spectral index 
$n$ as $\rho(r) \propto r^{-(9+3n)/(4+n)}$.

In reality the collapse of an initial overdensity is not so simple.
In particular, motions are not purely radial, and accretion does 
not happen in spherical shells (as assumed in the secondary infall 
model), but by aggregation of subclumps of matter which have already 
collapsed; a large fraction of observed galaxy clusters exhibit
significant substructure (Kriessler et al. 1995).
It is therefore very important to complement and compare 
these analytical studies with numerical simulations. 
These are not bound by such restrictions and so can tell 
if the gravitational collapse of a collisionless system eliminates  
all memory of the cosmological parameters which determined 
its initial conditions. They can also show whether scale-free universes,
which have no characteristic scale, give rise to scale-free 
power-law density profiles.
Work in this direction includes Quinn, Salmon \& Zurek (1986), 
Efstathiou et al. (1988) and West, Dekel \& Oemler (1987), but the 
somewhat conflicting results of these studies show that better 
numerical resolution is needed to settle the issue.
Recent results from higher resolution simulations (Navarro et al. 
1995 (hereafter NFW), Lemson 1995, Cole \& Lacey 1996 (hereafter CL), 
Xu 1996), 
produced by different $N$-body codes with different setups for the 
initial conditions, seem finally to agree on the following results:
\begin{enumerate}
\item
halo density profiles are curved, and are well approximated by a
fitting formula governed by a single {\em scale radius} $r_s$ and
belonging to the family of curves: 
$\rho(r) \propto x^{-\alpha}(1 + x^{\beta})^{-\gamma}$, $x = r/r_s$. 
NFW propose a model with $(\alpha,\beta,\gamma) = (1,1,2)$; another
candidate is the Hernquist (1990) (hereafter HER) profile: 
$(\alpha,\beta,\gamma) = (1,1,3)$.
\item
These fitting formulae provide a good model for halos formed 
in simulations of both scale-free and cold dark matter (CDM) universes.
\item
The value of the scale radius $r_s$ depends both on the initial
cosmology and on the mass of the halo in a way apparently related to
the formation time of the halos.
\end{enumerate}
Interestingly, a mass dependence for the scale radius was also
found observationally by Sanders \& Begeman (1994), who used
the HER model to fit the dark matter component when modelling
the rotation curves of a sample of spiral galaxies.

Despite the recent wealth of studies on this subject, the computational 
limits of present day machines are such that simulations of galaxy 
clusters are only now starting to reach a resolution sufficient to 
resolve reliably the dark matter structure of the central $\sim$ 100 kpc.
As a result several issues are still waiting for more detailed study.
Among them: have the results presented so far converged? That is, 
are they independent of numerical limitations? Does the trend of 
the scale radius with halo mass depend on numerical resolution? 
Are the simulated halos in dynamical equilibrium? And, especially,
do halo dynamics affects halos structure, for example, the shape of 
density profiles, and the scatter in the relation between the halo 
mass and the scale radius $r_s$?

More generally, the distribution of dark matter in the central regions 
of the halo is of particular interest; for example, the ability of a 
cluster to act as a gravitational lens, producing multiple magnified 
images of background galaxies, depends crucially 
on the mass content of the very central part (few tens of kpc) of the 
cluster, hence on the slope of the density profile at that scale.
Further questions to ask are then: what is the structure of 
cluster-size halos in the central few tens of kpc? Are results of
simulations compatible with recent lensing observations?

The purpose of the present paper is to address some of these points.
Section 2 presents the simulations. Section 3 is dedicated to
extensive numerical tests to establish the reliability of our results.
In Section 4 the halo formation process is interpreted using a simple
description in terms of its mass and {\em rms} velocity. Section 5 presents our
main results on halo density profiles, their dependence on the dynamical
configuration of the system, analytical fits to them, and the dependence
of these fits on halo mass. Section 6 discusses related topics,
like the dynamical equilibrium of halos, and the comparison of simulations
to dark matter observations from lensing studies. Finally 
Section 7 summarizes the results and presents some conclusions.

\section{The simulations}\label{sec:simul}

\subsection{Initial Conditions}\label{sec:ic}

To generate the initial conditions for a cluster simulation, we took 
a previously evolved cosmological simulation of a large region, 
a cube of side $L$, and selected a suitable cluster from this region. 
The initial conditions in the neighborhood of the cluster were then 
resampled with higher resolution in the following way.
All particles in the final cluster within a sphere of mean overdensity 
$\delta_0$, (typically $\delta_0 = 200$), were traced back to the 
unperturbed initial conditions, and the {\em Lagrangian} region 
$V_L$ containing them was enclosed in a cube of size $L^\prime<L$; in 
our case $L^\prime=0.25 L - 0.4 L$ was required. 
We then replaced the original particles in this cube with a larger 
number of lower mass particles, and perturbed them according to the 
same density fluctuations from the parent simulation, together with 
new fluctuations of higher frequency (up to the new Nyquist frequency), 
with amplitudes given by the theoretical power spectrum $P(k)$.
Since $V_L$ is irregular in shape, its volume is only a fraction 
of the volume ${L^\prime}^3$, typically 10\% to 15\% of it. To optimize
the use of the high resolution particles for the formation of the
cluster, we peeled off the high resolution cube and left high 
resolution particles only in an irregular {\em high resolution region}
that closely follows the shape of $V_L$, and has a volume roughly
twice that of $V_L$. 

The large-scale density and velocity field of the simulation was 
modeled as follows.
All particles in the original simulation falling outside the high 
resolution region (referred to as {\em background particles}) were
selected, and their mass and velocity were 
interpolated onto a spherical grid, using fixed angular resolution 
$d\theta = d\phi$, and with $dr = r d\theta$ in order to give 
approximately
cubic cells throughout the sphere. This operation reduces the number
of background particles to the minimum necessary to preserve the
large-scale tidal field of the original simulation.
Detailed tests have shown that a choice $d\theta = 7.5^\circ$, 
corresponding to $6000$ to $9000$ background particles, allows an 
accurate sampling of the tidal field. By construction, the mass 
of the background particles increases with distance from the high 
resolution region, and their gravitational softening $s$ was 
increased accordingly: $s \propto M^{1/3}$.
The sphere of background particles was always taken as big as the 
size $L$ of the parent simulation. The relevant parameter
here is the ratio between the maximum wavelength included in the
simulation and the size $L^\prime$ of the proto-cluster. Since 
in our case $L^\prime$ can be as big as $0.4 L$, the fluctuations 
mainly responsable for the initial density peak in the proto-cluster 
have wavelength $\lambda \approx 2 L^\prime - 3 L^\prime$ which is 
already of the order of the original box size $L$.
Therefore, taking a sphere smaller than $L$ would exclude important
contributions in the initial fluctuations and so would probably change
the final result, although we have not investigated along this 
direction.

\begin{figure*}
\epsfxsize=\hsize\epsffile{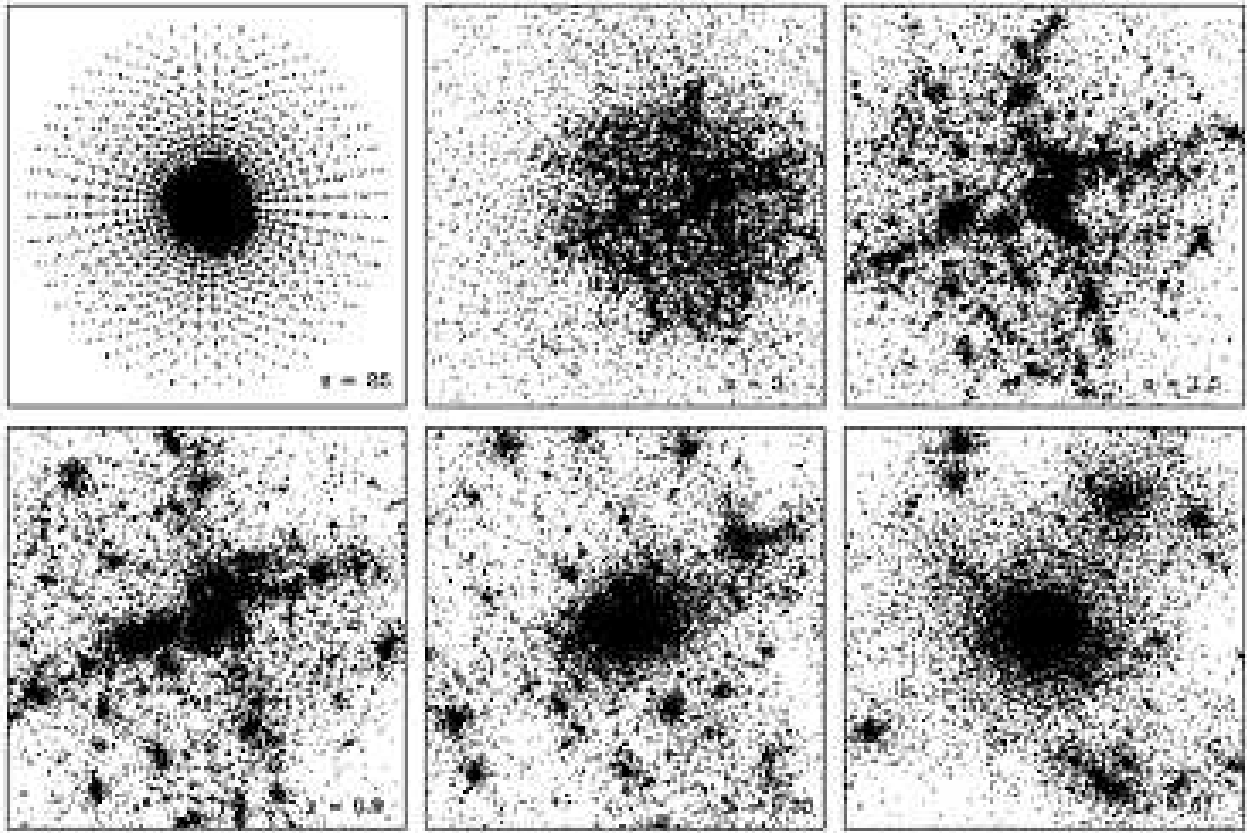}
\caption{Example of a simulation: time goes from left to right and 
from top to bottom. The first panel shows the initial conditions 
of the simulation, at $z=25$. The high resolution region in the 
centre is formed by $\approx 50000$ particles; the sphere of 8700 
background particles of varying mass provides accurate sampling
of the large scale tidal field.
The diameter of the simulation at $z=25$ is 6 Mpc in proper 
coordinates. In the following panels, only the high resolution 
particles are shown, at different redshifts. Each panel represents
a region of 12 Mpc (proper) on a side. The circle indicates the
virial radius $R_v$ of the densest clump of matter in the simulation
at that time. At $z=0$ there are roughly 24000 particles within $R_v$.}
\label{fig:ic}
\end{figure*}

The new initial conditions were finally traced back to a
higher redshift, such that the three-dimensional {\em rms} initial 
particle displacement in the high resolution region was less than
$0.75 \bar d$, with $\bar d$ the mean interparticle separation.
This ensures the validity of linear theory even in the more 
densely sampled regions.
By evolving from these conditions, the formation of the selected 
cluster can be followed with higher resolution than before, with 
a reasonably low computational cost.
An example of this setup, together with the simulation itself, is
shown in Figure~\ref{fig:ic}.

\subsection{Simulation details}\label{sec:numbers}

We started from a cosmological simulation of an Einstein-de Sitter 
universe, with scale free power spectrum $P(k) \propto k^n$, $n=-1$,
evolved using a Particle-Particle-Particle-Mesh code 
(Efstathiou et al. 1985) with $100^3$ particles in a $256^3$ grid
with periodic boundary conditions. The simulation box is 150 Mpc 
on a side, for a dimensionless Hubble parameter $h = 0.5$. 
All distances given in this paper are for $h=0.5$.
We selected the nine most massive clusters formed in the simulation
(containing typically a few thousand particles each), set up 
higher resolution initial conditions for all of them,
as described in the last Section, and carried out nine high 
resolution simulations.
For the evolution we used a tree-SPH code (Navarro \& White 1993), 
without gas, i.e. as a pure $N$-body code. This code has individual 
and arbitrary time stepping (Groom \& White 1996, in preparation).
The normalization for all simulations is such that at the final 
time the {\em rms} matter density fluctuation in spheres of radius 
$r = 8 h^{-1}$ Mpc is $\sigma_8 = 0.63$, in agreement with the 
observed abundance of local clusters (White et al. 1993).

The virial mass of the clusters, enclosing an average overdensity 
$\delta \rho / \rho = 178$, ranges from 
$M_v = 5.2 \times 10^{14} M_{\sun}$ to about 
$3 \times 10^{15} M_{\sun}$,
while their one dimensional {\em rms} velocity within the virial radius ranges
from 700 km s$^{-1}$ to 1300 km s$^{-1}$.
The average number of particles within the virial radius of each cluster
is $N_v \simeq 20000$.
The gravitational softening $s$ imposed on small scales follows a cubic
spline profile, and is kept fixed in physical coordinates. Its value
is $s=20 - 25$ kpc at the final time, depending on the
simulation. The typical maximum number of timesteps per simulation 
is of order 20000. Forces softened by a cubic spline can be roughly 
approximated by a Plummer softening $\epsilon \simeq s/1.4$.
The force resolution of our simulations is $L/s \simeq
6000$ to $7500$ for the box and of order of $R_v/s \simeq 100$
(with $R_v$ the virial radius) for single halos. The mass
resolution for one halo is, as we said, of order 20000.
The typical central density we resolve in the halos with the innermost
50 particles is of order of $10^6$ times the mean background density.

We can compare these parameters with those of recent simulations
which have addressed the same issue of the structure of 
dark matter halos in different cosmological models. We will quote
results for the most massive halos in each case.

The scale free $n=-1$ simulation run by Crone et al. (1994) 
have a force resolution $L/1.4 \epsilon = 730$ on the simulation 
box, and of about $R_v/1.4\epsilon = 30$ for each halo. Their 
mass resolution is of order of 6000 for the same objects.

The standard CDM simulation by Jing et al. (1994) has force resolution 
of about $3000$ for the box and 30 for massive halos, and a mass 
resolution of order 1000 particles for a typical massive halo.
They used 400 to 800 timesteps to evolve the system to the present
time.

The force resolution in the CDM simulations by NFW
is $\approx 12000$ for the box, and like ours ($\sim 100$) for 
single objects. Their mass resolution is 5000 to 10000 particles
per halo. Their typical maximum number of timesteps per simulation
is between 10000 and 100000.

Finally, the scale free $n=-1$ simulation performed by CL have a 
force resolution of about 2700 on the box, and of about 
40 on their most massive halos. Their mass resolution on these is 
about 7000 particles, and they evolved the simulation in $\approx
500$ equal timesteps. 

\begin{figure*}
\epsfxsize=\hsize\epsffile{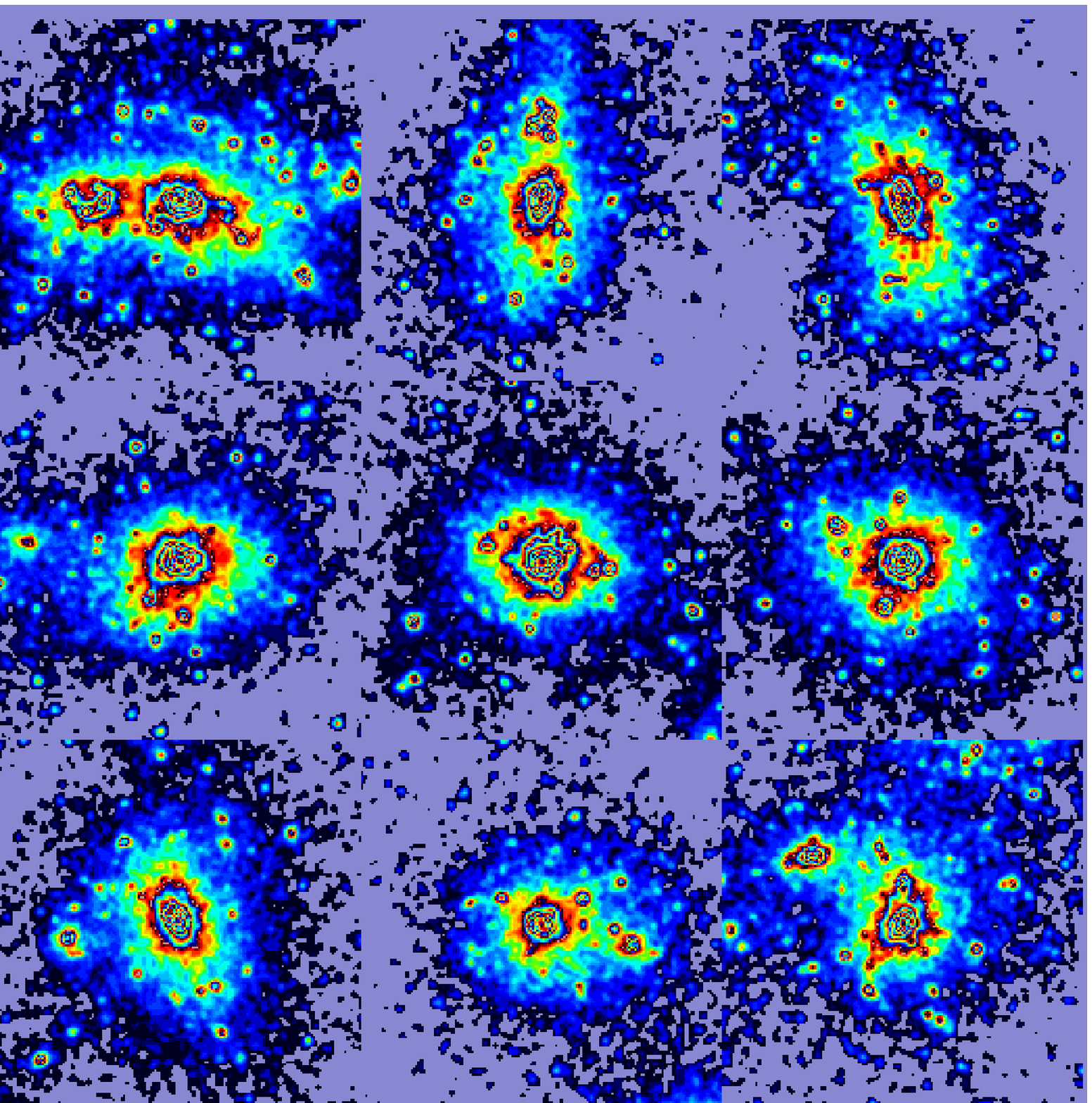}
\caption{Projected dark matter density of the nine halos presented 
in this paper. Images are shown from left to right and from top to 
bottom following the order of Table 1. All outputs at $z=0$; the 
size of each image is $3 R_v$ on a side, with $R_v$ the virial radius
of the halo. }
\label{fig:images}
\end{figure*}

This comparison shows that our simulations are the most accurate 
run so far, and we hope they will serve as a reference point for 
further studies. They took on average 200 hours of cpu time each, 
running on fast workstations.
For seven out of nine simulations we have 25 outputs, equally spaced 
in time from $t=t_i$ to the present time $t=t_0$. For the other two
we have fewer outputs.
The main physical and numerical parameters of the simulations are 
summarized in Table~\ref{tbl:1}.
Figure~\ref{fig:images} shows, for all halos, the projected dark 
matter density at $z=0$.
As we will show in Section~\ref{sec:tests}, we trust the results of
our density profiles down to a distance from the cluster centre equal 
to the gravitational softening radius. This gives us two order of 
magnitudes of spatial resolution for the clusters here presented.

\begin{table*}

\caption{Main numerical and physical parameters of the nine simulated
halos. The column content is: halo coding label; redshift of initial 
conditions; gravitational softening at $z=0$; number of particles
within the virial radius; virial mass; virial radius; one dimensional 
{\em rms} velocity within $R_v$; ratio of softening and virial radius;
number of minimum timesteps taken by the simulation.}

\begin{tabular}{crrrrrrrr}

Label & \multicolumn{1}{c}{$z_i$} & \multicolumn{1}{c}{$s$} &
\multicolumn{1}{c}{$N_v$} & \multicolumn{1}{c}{$M_v$}  &
\multicolumn{1}{c}{$R_v$} & \multicolumn{1}{c}{$v_{rms,v}$} & 
\multicolumn{1}{c}{$s/R_v$} & \multicolumn{1}{c}{$t_H/dt_{min}$} \\
  &   &\multicolumn{1}{c}{[kpc]}&  & \multicolumn{1}{c}{$[M_{\sun}]$} &
\multicolumn{1}{c}{[kpc]}   & \multicolumn{1}{c}{[km s$^{-1}]$} &  &  \\
\hline
 g15  & 17.   & 25  & 39400 & $2.99 \times 10^{15}$ & 3870  & 1260 & 0.0065 & 19000\\
 g23  & 25.   & 25  & 17400 & $6.76 \times 10^{14}$ & 2350  &  750 & 0.011 & 15000\\  
 g36  & 16.   & 25  & 18200 & $1.51 \times 10^{15}$ & 3070  & 1000 & 0.008 & 14000\\ 
 g40  & 22.   & 25  & 21300 & $5.32 \times 10^{14}$ & 2170  &  730 & 0.012 & 17000\\ 
 g51  & 26.   & 25  & 23500 & $1.38 \times 10^{15}$ & 2990  & 1000 & 0.008 & 19000\\ 
 g57  & 25.   & 20  & 24400 & $7.01 \times 10^{14}$ & 2380  &  780 & 0.008 & 21000\\ 
 g66  & 19.   & 25  & 21400 & $1.10 \times 10^{15}$ & 2770  &  920 & 0.009 & 21000\\ 
 g81  & 25.   & 25  & 14400 & $7.05 \times 10^{14}$ & 2390  &  750 & 0.011 & 15000\\ 
 g87  & 22.   & 20  & 16200 & $6.21 \times 10^{14}$ & 2290  &  740 & 0.009 & 21000\\ 
\hline

\end{tabular}
\label{tbl:1}
\end{table*}

\section{Cluster profiles: definitions and numerical tests}\label{sec:tests}

We want to investigate the distribution of dark matter in cluster-size
halos, in particular in their central regions. We will do this mainly
by showing various profiles (density, mass, velocity) obtained from
the simulations.

\subsection{Profile construction}

We defined profiles by binning the particles in spherical shells
centred on the cluster centre, as identified by the procedure described
in the next Subsection.

We tried both mass weighted profiles (with shells containing a fixed 
number $k=25$ or $k=50$ particles), and equally spaced logarithmic 
intervals of width $0.1$.
Since the profiles binned in equal logarithmic intervals are 
less noisy than the mass weighted profiles, and still let one describe
the halo properties down to small radii, we use them in the figures of
this paper. However, we performed some of the analysis also with mass
weighted bins, and did not find significant differences in the results.

The velocity profiles were calculated as follows.
First we defined the velocity reference frame by computing the mean
velocity of the cluster. Such a velocity was obtained by averaging 
over all particles within the radius enclosing half the virial mass.
We did not take all the particles in the cluster in order to exclude 
possible merging substructure, which would bias the mean velocity estimate.
This mean cluster velocity was then subtracted from the velocity of
each particle. The {\em rms} velocity within $r$ and the velocity
dispersions at $r$ were computed in this reference system, that is we
assumed a static non-rotating halo.

\subsection {Center definition}

The cluster centre was identified as the densest point of the dark matter
lump, found iteratively as the centre of mass of spheres of increasingly 
smaller radius. We started with a radius enclosing the whole system,
and stopped the iteration when a (fixed) small number of particles,
say $m$, were left in the sphere. The position of the centre of mass
of these $m$ particles was then identified as the final cluster centre.
The center found with this procedure corresponds with very good 
accuracy to the highest density peak of the simulation, even in 
cases of ongoing merging between different lumps, and also to the 
position of the particle with most negative gravitational potential 
(Lemson 1995, Tormen 1996 in preparation). The only free parameter 
of this method is the number $m$ that determines when the iterative 
calculation should stop.

Using a few clusters in our sample, we tested that the scatter in 
the centre location due to different choices of $m$ is negligible. 
In fact, the distribution of centre positions obtained by varying $m$ 
from 1 to 100 has a dispersion of about 5 kpc in the tested clusters,
well below the gravitational softening of the simulations.
We also found that the uncertainty in the centre position leads to 
differences in the profiles which are perfectly compatible with 
the Poissonian noise in the shells.
We thus defined the the centre of each of the nine clusters as the 
average between the centre positions found by varying $m$ from 1 to 100.

\subsection{Time integration accuracy}

A crucial ingredient for a reliable result, especially in the densest
region of the clusters, is the accuracy of the time integrator.
The use of a tree code with individual and arbitrary time steps is 
therefore ideal to follow with high accuracy the trajectory of particles 
travelling in very dense regions.
The time step of each particle is set by requiring that the relative
change of position and velocity of the particle be smaller than a
fixed tolerance. In our standard simulations, we chose a time integration 
tolerance corresponding to about 15000 to 22000 time steps per Hubble 
time, for the minimum time step actually taken by the particles. 
The minimum time step allowed in principle was at least ten times smaller
than this. We also set a limit on the maximum allowed time step which
corresponds to 24 time steps per Hubble time. Such big jumps were taken 
by a few background particles at large distance from the central system.

To make sure that such accuracy is high enough to properly resolve
the particles' orbit down to the cluster centre, we ran the last
$10^9$ years of a simulation three times, each time with different 
accuracy in the time integration scheme. The minimum timestep in the
three cases corresponded to $100000$, $30000$ and $9000$ timesteps 
per Hubble time. The halo had $N_v= 30000$ particles and a
gravitational softening of 36 kpc.
At the final time, we compared the profiles of the three runs: 
density, circular velocity, radial velocity dispersion and cumulative 
{\em rms} velocity, binned in logarithmic intervals 0.1 wide. We found 
very good agreement in all cases.
Small differences in the central region of the cluster were at 
about $1\sigma$ level, with $\sigma$ given by the shot noise in the 
bins at all radii down to the gravitational softening.
We conclude from this test that the standard tolerance used for our 
simulations allows a tracing of the particles' trajectory that is 
accurate enough for resolving the structure of the clusters even 
in the densest part.

\subsection {Gravitational softening and particle number}

The gravitational softening parameter $s$ and the number of particles 
composing the dark matter halo within the virial radius, $N_v$, 
are two closely related parameters.
For a given $N_v$, $s$ must be chosen big enough to maintain the 
collisionless evolution of the system, i.e. to avoid two-body relaxation 
phenomena, but small enough to reach the highest possible resolution, 
if one ignores cpu time limits. If one instead first fixes a value 
for the softening parameter $s$, then a sensible choice for $N_v$ is
the minimum number allowing one to resolve the central regions without
suffering two-body relaxation.

Several questions related to this point can be formulated.
The first is: what is the maximum scale $r_{soft}$ at which effects 
due to the softened gravity are visible or are important in the profiles?
Second: what is the minimum number of particles $N(r_{soft})$ one 
needs to have within such scale in order to properly resolve the halo 
structure down to $r_{soft}$, i.e. in order to make two-body relaxation 
effects negligible within the evolution time?
And what is the minimum number of 
particles $N_v$ one needs to have within the virial radius of the halo 
in order to have $N(r_{soft})$ particles within $r_{soft}$?
In this Section we will try to answer these questions by performing
test simulations with different values for $s$ and for $N_v$.
We will show results from two orthogonal tests: one fixing the
particle number and varying the softening parameter, the other fixing
the softening and varying the number of particles.

\subsubsection{Test of gravitational softening}\label{sec:soft}

Let us address the first question posed above: how big is the
effect of the force softening on the results. In order to answer this
question we run different simulations of the same cluster, scanning 
the parameter space ($s$,$N_v$) along curves at fixed $N_v$.
We run our test on cluster g57, the one that looks more in equilibrium
both from visual inspection and from the criteria that will be given
in Section~\ref{sec:evol}.
We ran 4 simulations: three have the same particle number $N$, which
gives $N_v \approx 24000$ at the final time,
but differ in the value of the gravitational softening parameter.
This was taken to be $s = 20$, $50$ and $100$ kpc respectively,
where the $20$ kpc run refers to the standard simulation.
We also ran a larger simulation of the same cluster, with 
$N_v \approx 38000$ and $s=10$ kpc, which took 32000 minimum 
timesteps. We will use this simulation as a {\em reference} 
for the all numerical tests presented in this Section. That is, we 
will quantify the effect of varying $s$ and $N_v$ by comparison to 
this {\em reference} simulation.

It is possible that Poissonian fluctuations in the particle counts, 
or transient phenomena like a subclump crossing the cluster centre,
cause differences in the halo profiles. 
In the present test, such differences would be spurious ones, because 
they are not strictly due to the change of the softening parameter, 
and we would like to remove them from the results.
Since we do not have a large number of simulations to average on, we
average instead over the last few snapshots for each of the four cluster
runs, under certain hypotheses of regularity.
Specifically, we select, for each cluster, all the time outputs that at 
the same time satisfy the following two characteristics:
the cluster virial mass be at least 90\% of the final mass;
the redshift of the cluster be less than 0.1.
The first requirement minimizes possible dependence of the results
on the cluster mass; the second minimizes possible evolutionary effects.
In practice this means selecting the last four outputs of each simulation. 

The average is performed on equally spaced decimal logarithmic
intervals of width 0.1 in $\log(r)$.
In this way we obtain for each cluster some {\em average} profiles, 
which we consider as representative of the {\em typical} dynamical 
configuration of the halo at recent times.

\begin{figure}
\epsfxsize=\hsize\epsffile{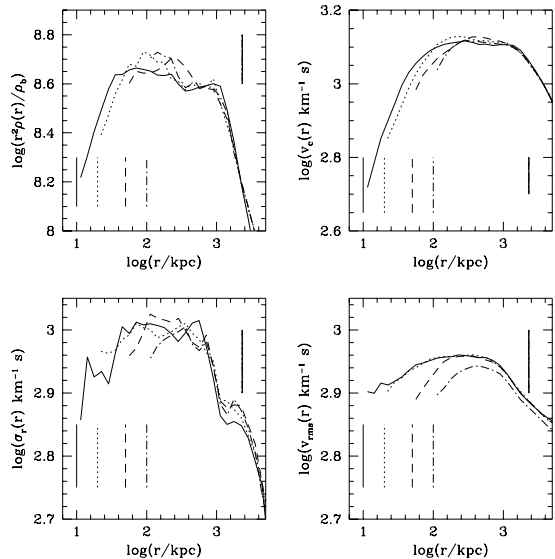}
\caption{Test of gravitational softening. The solid curve refers to
the reference simulation: $N_v = 38000$, $s=10$ kpc. The other curves
refer to halos with $N_v \approx 24000$ and $s=20$ kpc (dotted),
$s=50$ kpc (dashed) and $s=100$ kpc (dot-dashed), The vertical bars 
on the left indicate the value of the gravitational softening for each
profile. Those on the right indicate the virial radius.}
\label{fig:soft}
\end{figure}

The radial profiles for density $r^2\rho(r)$, circular velocity $v_c(r)$, 
shell radial velocity dispersion $\sigma_r(r)$ and one dimensional {\em rms} 
velocity in spheres $v_{rms}(r)$ are shown in Figure~\ref{fig:soft}.
The solid line indicates the profile of the {\em reference} simulation
($s=10$ kpc). The profiles are plotted down to a radius equal to
the gravitational softening, indicated by the vertical lines on 
the left, provided that there are at least 50 particles within 
the corresponding bin.
The virial radius is given by the vertical lines on the right.

Simulations with softer forces are expected to produce halos 
that are less dense in the centre and have lower velocities.
This is clearly the trend in the figure. However, differences in
the profiles at radii larger than the softening radius are of the
order of 20\% or less in all cases. 
This result answers the first question we posed above: we can 
trust the profiles of our standard simulations to $\sim 20\%$ or better 
down to a distance from the centre $r_{soft}$ equal to the
gravitational softening $s$. This is quite a good accuracy, being 
comparable to the noise level of mass weighed profiles with 25 
particles per bin, and recalling that density in halos varies over
5 orders of magnitude.

\subsubsection{Test on particle number}

Having found that $r_{soft} \simeq s$, we may rephrase the second 
question posed at the beginning of this Section as follows: 
how many particles are needed within a softening radius from 
the cluster centre in order to trust the density profiles down 
to $r = s$ and to avoid two-body relaxation effects?

To answer, we made another test, which is orthogonal to that 
just described. We run different simulations of the same cluster
using the same gravitational softening $s$, but changing the 
particle number $N$.
We hope this will give us some idea of the minimum number of
particles needed to resolve the cluster down to a given scale
from the centre.
This set of simulations thus scans the parameter space ($s$,$N_v$) along 
curves at fixed $s$. We chose $s=20$ kpc, and particle numbers roughly
corresponding to 24000, 12000 and 7000 particles within the
virial radius of the halo. For this test we used the same cluster
selected above, i.e. g57.
We averaged the profiles from the last four time outputs, in the same 
way described above, to isolate the effect of particle number from 
transient dynamical effects.

\begin{figure}
\epsfxsize=\hsize\epsffile{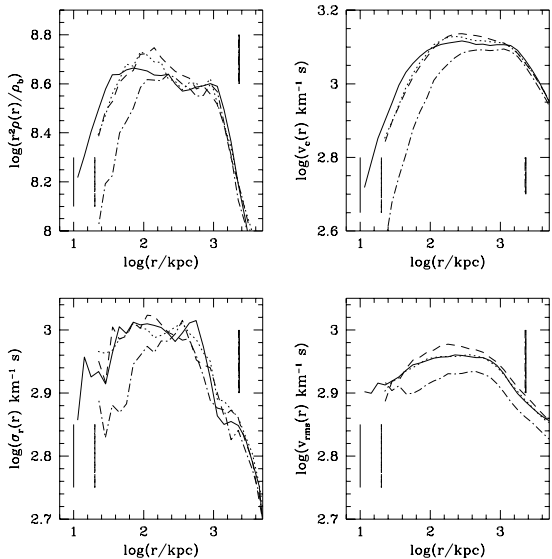}
\caption{Test on particle number. The solid curves refer to the
reference simulation: $N_v = 38000$, $s=10$ kpc. The other curves
have all $s=20$ kpc and $N_v = 24000$ (dotted), 12000 (dashed) and
7000 (dot-dashed). The vertical bars indicate the softening and
virial radius for all simulations.}
\label{fig:npart}
\end{figure}

Figure~\ref{fig:npart} shows the resulting profiles. 
The simulations with 24000
and 12000 particles within the virial radius agree quite well with
the reference run: the differences are less or of order 20\% in the
profiles. The run with $N_v \approx 7000$ instead has profiles 
systematically lower than the reference ones; differences at small
radii are more than a factor of two in the density, of around 50\%
in the mass and of order 30\% in the radial velocity dispersion. 
The least affected profile is the mean {\em rms} velocity, which differs 
less than 20\% from the reference value. 
It is only at radii larger than about 100 kpc,corresponding to 
a sphere enclosing 250 to 300 particles, that these profiles agree
with those of the other simulations.

Note that the simulation with $N_v \approx 7000$ has only 
$\approx 10$ particles within $r = s$, so there are not enough
particles to reliably define the profiles down to $r = s$. 
We conclude that, in order to achieve a dynamical resolution of about 
$0.01 R_v$, or -2 in log$(r/R_v)$, to better than 20\% in the
profiles down to the softening radius, one needs at least 
$\approx 12000$ particles within the virial radius. 
This of course for the kind of models analyzed here. 
Equivalently, we can say that 10 particles within a radius equal to 
the softening radius are not enough to avoid discreteness effects 
in the central 100 kpc, for a cubic spline softening.

In order to confirm and extend this result we ran other sets of simulations,
each set with a fixed gravitational softening and different particle 
numbers. In one case we took $s = 36$ kpc, corresponding to 
log$(s/R_v) \simeq -1.8$, and $N_v \approx$ 30000, 13000, 7000 and 
3000. 
In a second set we took $s = 80$ kpc, corresponding to 
$\log(s/R_v) \simeq -1.5$, and $N_v \approx$ 2,500, 1,200 and 800. 
We also ran a simulation with $s = 10$ kpc or $\log(s/R_v) \simeq -2.3$
and $N_v = 24000$ to compare to the reference simulation.
Looking at the profiles resulting from all these tests we found that, 
in order to produce reliable profiles down to $r = s$ one needs at 
least $N_v \approx 38000$ for $\log(s/R_v) = -2.3$, and 
$N_v \approx 7000$ for $\log(s/R_v) = -1.8$.
With $N_v$ smaller than these values, the profiles exhibit the same trend:
a flattening of the density in the centre and lower velocities. The
same profiles agree however with the reference ones at radii enclosing
more than 200 to 250 particles.
On the other hand, $N_v \approx 800$ seems enough to resolve
scales down to $\log(s/R_v) = -1.5$.

\subsection {Numerical tests: Conclusion}

We can summarize the results of our tests as follows.
\begin{enumerate}
\item 
The uncertainty in the halo centre is well below the value of
the gravitational softening, and the corresponging errors introduced 
in the profiles are below the shot noise level.
\item 
The time integration of the simulations is performed with 
sufficient accuracy to ensure the correct orbit calculation for all
particles, including those in the densest regions.
\item 
The softening of gravity at small scales causes appreciable
effects only at scales below the softening $s$ itself. At scales larger
than $s$ the profiles are affected by softening by less than 20\%
in all cases.
\item We find that, for a gravitational softening $s \simeq 0.01 R_v$,
the number of particles required to avoid discreteness
effects is conservatively $N_v~\ga~10000$. Smaller numbers of particles 
cause a flattening of the density profile in the centre and significant 
differences in the results out to a radius enclosing $\approx 250$ 
particles.
\end{enumerate}

\section {Cluster formation process}

We restricted our study to a specific {\em fiducial} cosmological
model, namely an Einstein-De Sitter universe with zero cosmological 
constant and scale-free density perturbation spectrum $P(k)$ with a 
spectral index $n = -1$.
This slope is similar to that of the standard CDM 
power spectrum on cluster scales. Therefore, the conclusions that 
we will draw from this work should also give us some insight 
on the formation of galaxy clusters in CDM universes.

The aim of this paper is not to compare different cosmological models,
but rather to study with high accuracy the cluster formation process 
in a specific but representative case. For this we have many independent 
examples of well resolved
halos coming from the same universe and from a relatively limited mass
range. We can therefore estimate the expected scatter in results,
for the evolution of clusters that are in principle very similar.
In particular we can study the dynamics of the cluster growth, which
can range from a quasi-static accretion state to violent merging events.
We can also see how these states alternate during cluster evolution, 
and how they influence cluster dynamics.

\subsection{Evolution of mass and {\em rms} velocity}\label{sec:evol}

\begin{figure}
\epsfxsize=\hsize\epsffile{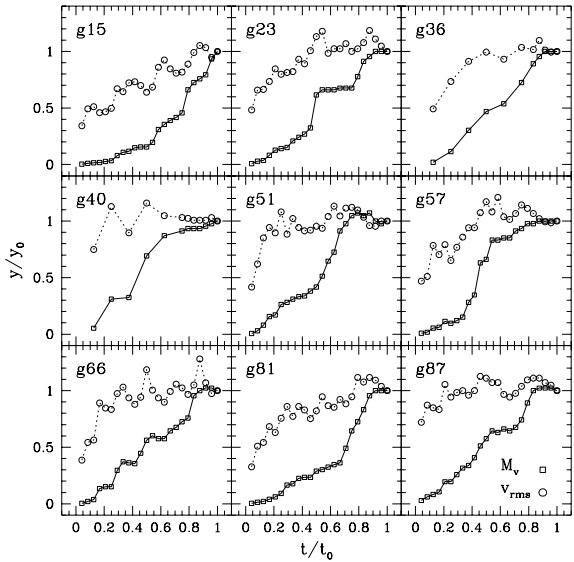}
\caption{Evolution of virial mass and {\em rms} velocity. Quantities are
normalized to their value at $z=0$.}
\label{fig:evol}
\end{figure}

Figure~\ref{fig:evol} shows, for all objects in our sample, the time 
dependence of the total (i.e. virial) cluster mass and of the 
one-dimensional {\em rms} velocity within the virial radius. 
At early times the most massive cluster progenitor was taken.
Each panel refers to a cluster. The quantities are measured at 24 
different times, except for clusters g36 and g40, for which we have 
stored fewer data.

One can easily see from the trend of the virial mass that the 
cluster accretion process is characterized by two quite different 
phases. The first is recognized by sudden leaps in the mass curve,
which correspond to {\em merging} events, i.e. to rather massive
lumps crossing the virial radius.
The number of merging events during the simulation is three to 
five in most cases. The mass of the merged substructure, roughly 
estimated by the leaps in the mass curve, can range from about 20\% 
of the mass of the main lump to a value comparable to it.
The second phase is a {\em relaxation} phase, which follows each merging 
event, and corresponds to periods during which the cluster does not
significantly accrete, but rather processes the infallen matter. 
During this period, the substructures orbit within the virial radius, 
swing back and forth through the cluster centre and are gradually 
destroyed. The mass versus time curve is thus characterised, in this 
phase, by a plateau or by gentle growth.

Even in absence of accretion, the virial Mass of a cluster will grow, 
since the expansion of the universe makes the background density 
decrease and the nominal boundary of the cluster thus expands. 
A virial mass that remains constant or even decreases with time, as 
is sometimes observed in the plots, indicates that there is a net
outflow of matter at the virial radius.

The alternation of the {\em merging} and {\em relaxation} phases is
typical of all the clusters in our sample, although the length of 
time a cluster spends in one phase or the other varies, as can be 
inferred from Figure~\ref{fig:evol}. 

Looking now at the behaviour of the average {\em rms} velocity of
particles within the virial radius, we see that the merging phase 
causes an increase of $v_{rms}$. This is due to the kinetic energy 
acquired by the subclump as it falls in the potential well of the 
main cluster. The {\em rms} velocity peak corresponds to the first passage 
of the subclump through the cluster centre, so it is slightly delayed 
with respect to the jump in virial mass. This finding agrees well with
the results of Crone \& Geller (1995).

After this first passage the {\em rms} velocity gradually decreases 
towards a minimum, and the cluster relaxes while 
it is destroying the recent acquisition.
The peak value is usually $\approx 15\%$ higher than the following 
equilibrium value, but in some cases it is 30\% higher.

The relaxation of the cluster implies a decrease of the {\em rms} velocity,
whether or not there is significant substructure.
In fact we stress that, from the time sequence of outputs, it is evident 
that only the {\em first} passage of a merger through the centre 
of the main cluster produces a significant rise in the {\em rms} velocity. 
Subsequent passages are already damped and are much slower: they do
not show up in the curves.
In the same way, when a merging clump has already crossed the cluster 
border but has not yet aquired significant kinetic energy from its 
fall, the {\em rms} velocity is not affected by its presence.
So the presence of substructure does not, in itself, change the 
velocity distribution of the cluster, and a very lumpy cluster can
be still {\em relaxed} from the velocity point of view. 
In this sense, clusters in an $\Omega = 1$ universe, which are 
accreting up to the present day, may also go through phases of 
relaxation.

We may ask several questions at this point: can we show
more quantitatively that this alternation of states really 
corresponds to changes in the system equilibrium? If so, how
much is the system affected in terms of, e.g., mass estimates
based on the velocity dispersion? And finally, what fraction of its
evolution does the system spend, on average, in a relaxed or 
perturbed state? We will discuss these equilibrium issues in
Section~\ref{sec:dyneq}, after analysis of the halo profiles.

\section {Halo profiles}

In this Section we present our results for the profiles of the nine 
halos drawn from our $N$-body simulations.
There has been much debate over the {\em expected} shape of dark matter
density profiles, the link of this to the initial cosmology, and 
the astrophysical implications of the existence of a core
in dark matter halos.
Here we limit our study to one model, an Einstein-De Sitter 
universe with zero cosmological constant and a scale-free density 
perturbation spectrum with spectral index $n = -1$. 
However, we can test to what extent cluster-size halos in a 
representative model like this are compatible with the observations.

\subsection {The profiles}\label{sec:profdata}

We can think of two kinds of approaches: an everage study, where 
different halos (or different outputs from the same halo) are averaged 
to obtain some mean cluster properties, or a case by case comparison,
where individual outputs from the simulations are considered in order
to have an idea of how and how much the various profiles change when 
the clusters evolve through different dynamical states.
For this reason, we will present results for the {\em average}, 
{\em most relaxed} and {\em most perturbed} configuration of each 
cluster. By average configuration we mean the one defined in
the Section on numerical tests. The most relaxed and most
perturbed configuration were defined using the curves of evolution 
of the virial mass and {\em rms} velocity in the following way.
To select the most relaxed output, we looked at Figure~\ref{fig:evol}
and chose for each halo the most evolved output situated at the end 
of a series showing at the same time a flat or smoothly increasing 
mass curve and an overall decrease of the {\em rms} velocity.
Since the halos in our sample usually show a relaxation phase 
towards the end of their evolution, the snapshots thus selected 
usually correspond to the most evolved time ($z=0$) or to the one 
immediately preceding it ($z\simeq 0.01$). Two exception are 
cluster g51, for which we selected the output at $z = 0.08$ and 
cluster g15, for which we chose the output at $z=0.19$. 
Conversely, we chose the most perturbed output as the one 
corresponding to a maximum in the velocity dispersion curve,
following a recent merging. Here also we chose the most evolved 
among the possible outputs, to retain the maximum mass resolution.

Averaging over different snapshots of a cluster allows us to 
reduce somewhat the noise in the profiles.
Since this procedure is in some sense equivalent to averaging 
over different clusters at the same output time, the average
over different dynamical configurations of the same object 
will hopefully produce data which are more representative 
of the most likely state of a cluster, and will perhaps 
allow a better comparison with the observations.
After all, real clusters are observed in various dynamical 
states, and not in their most relaxed configurations.
On the other hand, the relaxed profiles are perhaps closer 
to our idealized theoretical models, since their dynamics
should be less perturbed and more similar to a simple spherical
system. Therefore one hopes that the relaxed shape of halos
is more similar to the analytical predictions.
We recall however that the configurations we call {\em relaxed} 
do not correspond to true static systems, because they
usually still have a lot of substructure; accretion never stops 
in an $\Omega = 1$ universe.

\begin{figure}
\epsfxsize=\hsize\epsffile{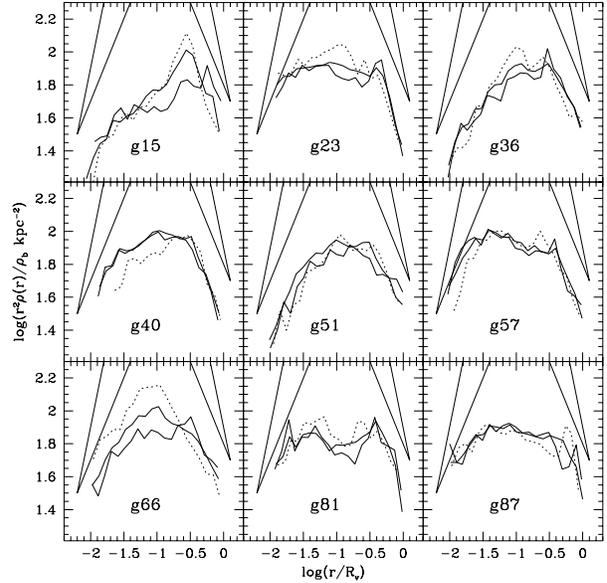}
\caption{Density profiles of our nine halos. For each halo we plotted 
three profiles, corresponding to the {\em average} (AVE: thick solid line), 
{\em most relaxed} (REL: thin solid line) and {\em most perturbed}
(PER: dotted line) dynamical configuration. Straight lines correspond to 
asymptotic logarithmic slopes $-1$ and $0$ at small radii and $-3$ and $-4$ 
at large radii.}
\label{fig:dendata}
\end{figure}

On the other hand, perturbed configurations may present the
cluster under extreme non-equilibrium conditions, which 
are nevertheless useful to give some idea of the range of
configurations one may happen to observe in the same
object. Figure~\ref{fig:dendata} shows the density profiles, 
all binned in logarithmic intervals of width 0.1.

On average, the density profiles of the three kinds show the following
features: a steeper logarithmic slope at large radii, an intermediate 
region where the slope bends towards the singular isothermal sphere 
value $-2$, and finally an inner region where the profile becomes 
gradually flatter. Sometimes at intermediate radii the profile 
follows almost a power law; sometimes the change of slope is more 
gradual and relatively smooth. 
The spikes visible in many clusters at large radii correspond
to merging objects which have just crossed the cluster virial 
radius.

We see that, in general, the three profiles of each cluster 
are fairly similar. Only two clusters (g15 and g66) show 
significant differences, of the order of a factor of 2, 
between the averaged, relaxed and perturbed density profiles. 
The other halos have profiles which differ by roughly 20\% in 
the averaged and relaxed configurations, or 40\% including the 
perturbed outputs. This differences are not big, and the result
tells us that the halo density, once averaged in spherical shells,
is not a very sensitive measure of dynamical evolution.

\begin{figure}
\epsfxsize=\hsize\epsffile{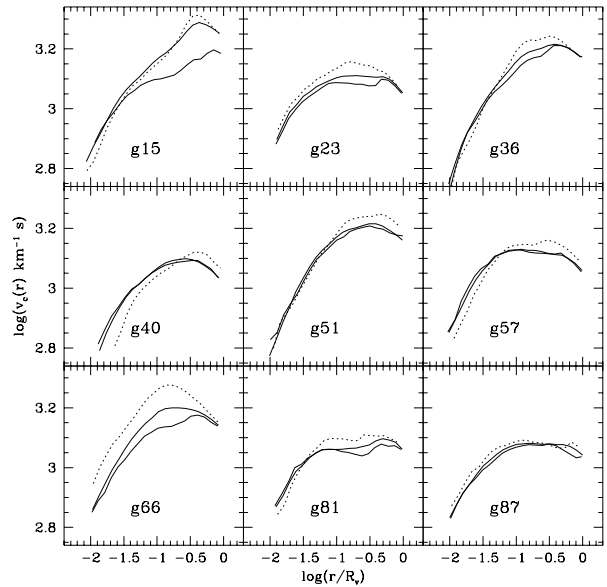}
\caption{Circular velocity profiles for the nine halos. Line types 
are as in Figure~\ref{fig:dendata}.}
\label{fig:vcdata}
\end{figure}

The same comparison is made for the circular velocity profiles 
in Figure~\ref{fig:vcdata}. 
Here the general trend is a rising circular velocity at small 
radii, a maximum and then a decrease at large radii. 
The departures from the isothermal behaviour $v_c = const$
are not due to limited resolution, but are real as was 
demonstrated in the Section dedicated to numerical tests.
In particular, the decrease at large radii always starts well
within the virial radius of the clusters.
Differences between the various configurations are smoother here, 
since their effect is integrated over the cluster. Again 
the averaged and most relaxed profile agree quite well, and also 
the perturbed outputs show average differences of roughly 20\%, 
but up to 60\% for the two cluster mentioned before.

\begin{figure}
\epsfxsize=\hsize\epsffile{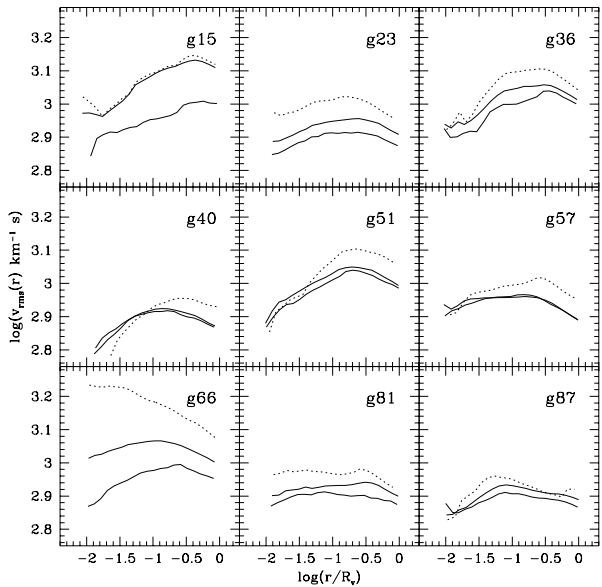}
\caption{One dimensional {\em rms} velocity within spheres of radius r, for
our halos. Lines are as in Figure~\ref{fig:dendata}.}
\label{fig:vrmsdata}
\end{figure}

The average one dimensional {\em rms} velocity within spheres of radius 
$r$, for the three configurations and for each cluster, are plotted 
in Figure~\ref{fig:vrmsdata}.
The profiles show an {\em rms} velocity that is usually increasing
with radius at small $r$, but sometimes is flat. After reaching
a maximum, $v_{rms}$ decreases slowly at large radii.
Cluster g66 exhibits a larger difference between the three
configurations because it shows a merger passing through its centre
in the perturbed and averaged profile.

Differences between the perturbed and relaxed configuration usually
are of the order of 15\% at the virial radius; in two cases they are
as big as 30\% to 40\%. A na\"{\i}ve application of the Virial Theorem 
would therefore miss the true mass of the halo by 30\% on average. 
We will come back to this point in more detail in Section~\ref{sec:dyneq}.

The radial velocity dispersion $\sigma_r(r)$ for the nine halos 
follows the behaviour of the {\em rms} velocity, rising with $r$ at 
small radii and decreasing at large radii. In this case the 
curves are more noisy and the curvature more pronounced because 
$\sigma_r(r)$ is computed for each shell while $v_{rms}$ is 
averaged on all particles within $r$.
The halos are thus not generally isothermal. However, departures 
from the isothermal model are not very big, at least for the 
averaged or relaxed outputs: the typical variation in  radial 
velocity dispersion over the range of radii resolved by the 
simulations ranges from about 45\% for the averaged and 
relaxed configuration to about 70\% for the most perturbed outputs.

\begin{figure}
\epsfxsize=\hsize\epsffile{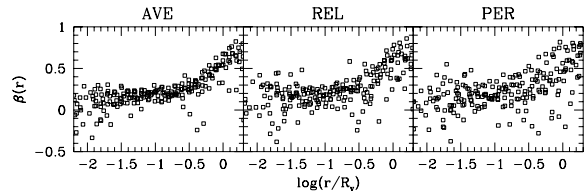}
\caption{velocity anisotropy parameter (defined in the text). In each panel
the profiles from all nine halos are plotted. Left panel refers to
averaged outputs, central panel to most relaxed outputs, right panel
to most perturbed outputs.}
\label{fig:beta}
\end{figure}

Lastly we show in Figure~\ref{fig:beta} the behaviour of the velocity
anisotropy parameter, $\beta(r) = 1 - \sigma_t^2(r)/2\sigma_r^2(r)$,
where $\sigma_r(r)$ and $\sigma_t(r)$ are the radial and tangential 
velocity dispersion, and the tangential velocity of each particle is 
defined by $\sigma_t^2 = v^2 - \sigma^2_r$. A value $\beta(r) = 1$ 
means purely radial orbits at that radius; an isotropic velocity 
field has $\beta(r) = 0$, while $\beta(r) <0$ means a predominance 
of tangential motions.

Each panel shows $\beta(r)$ for all halos together.
The left panel refers to the averaged outputs, the central to the
most relaxed ones, the right panel to the most perturbed
outputs. Although the scatter in the quantity is different, the trend 
is the same in all cases: $\beta(r) \simeq 0.2$ for $r \leq 0.2
R_v$, then it steadily increases with $r$, and reaches 
$\beta(r=R_v) \simeq 0.6$. Radial motion predominates at large
radii, while in the inner part of the halo orbits are more nearly 
isotropic. The transition between these two regimes happens at around
one fifth of the virial radius. In one case (halo g87) the velocity 
field is more isotropic also at large radii.

The results for $\sigma_r(r)$ and $\beta(r)$ are in good qualitative
agreement with those of Crone et al (1994) and Cole \& Lacey (1996),
although the value of $\beta(r)$ we find is slightly larger than theirs.
In their model of cluster A2218, Natarajan \& Kneib (1996) find a
negative $\beta(r)$ at small radii; although in contrast with our
average result, Figure~\ref{fig:beta} shows that $\beta<0$ can
occasionally be measured.

\subsection {Analytic fits to the profiles}

In this Section we present possible analytic fits to the profiles 
we have just shown.
We will show results of the analysis performed on the averaged 
profiles, as their features are similar to those of the most 
relaxed ones. We will however present the conclusions also
for the most relaxed outputs. We will not consider the perturbed 
profiles in what follows.

We should first ask what is the purpose of an analytic fit. 
Viewed as a data compressing technique, one may want to find a
fitting function that gives a good overall description of the 
observed profile at all radii, regardless of its true dynamical 
significance.
However, at a deeper level one might also hope that the analytic 
fits have some real dynamical meaning. In this paper we deal with 
the first approach only.

From Figures~\ref{fig:dendata} to \ref{fig:vrmsdata} it is clear
that the density and circular velocity profiles are not well 
represented by a single power law: they may exhibit a power-law
behaviour at some radii, but this does not extend throughout
the cluster. The departures from a single power law, at small 
and large radii, are real and are not due to numerical inaccuracies,
as we showed in Section~\ref{sec:tests}.
We will thus limit our comparison to curved fits. In particular,
we are going to consider the analytic profile recently proposed by
NFW and the HER profile.
Both describe the density profiles as a curve with a single 
{\em scale radius} that sets the scale of transition between 
two power laws. These profiles have been presented in the 
Introduction; we recall their form here:

NFW fit:
\be
\rho(x) = \rho_0 x_s^3{1\over x(x + x_s)^2};
\ee

HER fit:
\be
\rho(x) = \rho_0 x_s^4{1\over x(x + x_s)^3};
\ee
where $x=r/R_v$, and $x_s = r_s/R_v$, with $r_s$ the scale radius 
of the fit. The value of $\rho_0$ is determined by imposing that 
the mean density within $R_v$ is $178$ times the mean background 
density. The corresponding circular velocities are:

NFW:
\be
v_c^2(x) = 4\pi \rho_0 R_v^2 x_s^3 G 
\left[{1 \over x} \ln( 1 + {x \over x_s}) - {1 \over x + x_s} \right];
\ee

HER:
\be
v_c^2(x) = {GMx \over R_v(x + x_s)^2},
\ee
where $M$ is the total mass of the system.

Both profiles have the same asymptotic behaviour at small radii: 
$\rho(r) \propto r^{-1}$ as $r \to 0$.
The difference between them is at large radii, where the NFW profile
is shallower, with $\rho(r) \propto r^{-3}$ as $r \to \infty$. 
The HER profile instead behaves as $\rho(r) \propto r^{-4}$ 
at large radii.
In the range of interest, i.e. for $0.01 \la r/R_v \la 1$
the HER profile is the most curved of the two, because the
difference between the two power laws is larger.

We have fitted the profiles of our nine dark matter halos
with both analytic models. The best fit was determined using the 
logarithmic density profile. Because the circular velocity is an 
integrated estimate, the limits on resolution at small scale 
influence the profile at larger radii than is the case for the 
density. The minimization was performed by a standard Chi Square 
method in the range $s \sim 0.01 R_v \leq r \leq R_v$ (with $s$ 
the softening radius).
We used profiles binned in equally spaced logarithmic intervals 
of width 0.1, as described above. As a result, our fit is not mass 
weighted, but gives equal weight to small and to large radii.

\begin{figure}
\epsfxsize=\hsize\epsffile{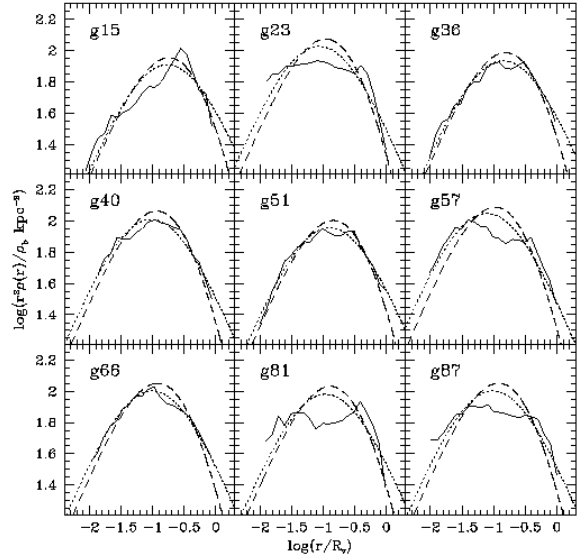}
\caption{The averaged (AVE) density profiles of the nine halos (thick
solid curves) are plotted with the best fitting analytical models. 
Dotted curves are NFW profiles, dashed curves are HER profiles.}
\label{fig:adenfit}
\end{figure}

\begin{figure}
\epsfxsize=\hsize\epsffile{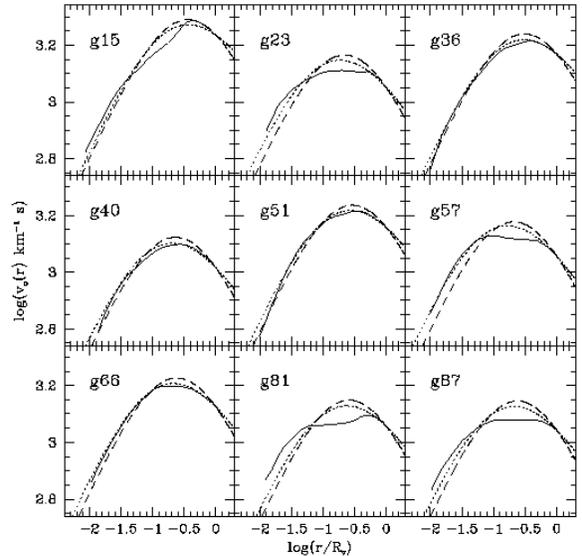}
\caption{Circular velocity profiles for the AVE configuration of our
simulations plotted with the best fitting NFW and HER profiles. Line
types are as in Figure~\ref{fig:adenfit}.}
\label{fig:avcfit}
\end{figure}

\begin{figure}
\epsfxsize=\hsize\epsffile{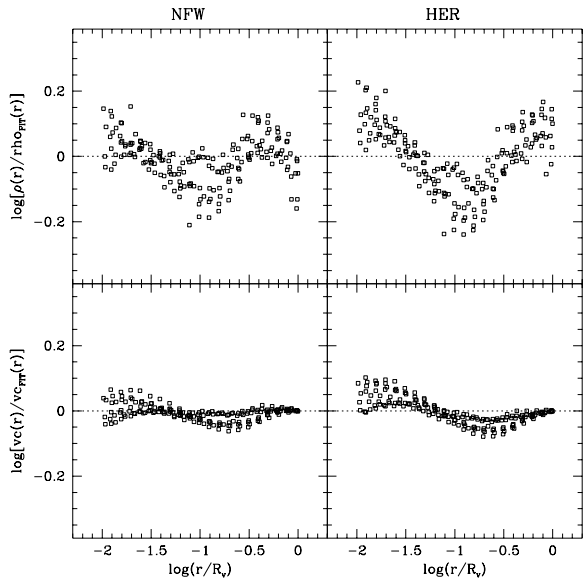}
\caption{Residuals of the fits to the AVE profiles of the previous
two figures.}
\label{fig:averes}
\end{figure}

Figures~\ref{fig:adenfit} and \ref{fig:avcfit} show, for the averaged 
configuration, the density and circular velocity profiles respectively, 
and their best NFW and HER analytical fit. 
The thick solid curve is the actual profile of the simulated halo, 
the dotted curve is the best fitting NFW profile, and the dashed curve 
is the best fitting HER profile. Recall that 
the fit has been made on the density profiles. Therefore in some 
cases the best-fitting curve for the density does not correspond 
to the best-fitting curve for the circular velocity.

Figure~\ref{fig:averes} shows a summary of the fitting results. 
It refers to the averaged profiles, and plots the residuals of both
the NFW and HER profiles for the density and circular velocity. 
One can see that on average the density profiles of the 
simulated clusters look steeper than both analytical fits at 
small radii, and shallower than the HER fit at large radii. 
Part of the positive residuals at large radii may be due to the 
presence of merging subclumps.
The scatter ($1\sigma$ dispersion) in the residuals is smaller 
for the NFW profile, and larger for the HER profile, 17\% and 26\%
in density, respectively.
The circular velocity is fitted with smaller scatter,
6\% and 10\% for NFW and HER respectively. If we consider the
fact that the curves extend over two orders of magnitude in radius,
and over four orders of magnitude in density, both fits can be 
considered quite good. The NFW model fits our simulations on
average 50\% better than the HER model, down to at least 
$r \simeq 0.01 R_v$.
Since the simulated profiles are in general less curved than either 
of the models at intermediate radii, the maximum circular velocity 
estimated by the fits is generally bigger than the true one, although 
the difference is small.

One can repeat the same kind of analysis on the most relaxed 
outputs of each halo. The corresponding residuals have however 
a larger scatter: the $1\sigma$ dispersion in their distribution 
is 24\% and 34\% for the density fits, for the NFW and HER models 
respectively, and 10\% and 12\% for the circular velocity fits. 
Therefore the averaged profiles turn out to be slightly better
fitted by the analytic curves than the relaxed profiles.
We are somewhat surprised by this result, and would have expected 
the converse. It may be that the bigger intrinsic noise 
in the relaxed profiles (which are built from only one output)
is responsible for the bigger deviations.
At any rate, differences between the averaged and most relaxed 
profiles are small.

\subsection{In search of equilibrium}\label{sec:evo}

We want to push the equilibrium issue a bit further. That is,
we would like to know how the halo characteristics would
change with respect to those measured so far, if the system
reached a real dynamical equilibrium, and all substructure
was processed and destroyed.
Since this cannot happen in an Einstein-de Sitter universe, 
where galaxy clusters are dynamically young and accretion
goes on for ever, we have to break the clustering hierarchy 
in some way, and then let the halos relax.

For this purpose we made the following experiment.
We chose the last output of each simulation, corresponding to
the present time, and removed from it all the particles beyond
a fixed radius $R_{cut}$ from the cluster centre. We choose
$R_{cut} \simeq 2 R_v$. We then evolved the remaining particles,
that is those forming the halo out to $R_{cut}$, with void boundary 
conditions around it, for another Hubble time. 
By cutting the halos at twice their virial radius we allowed some
additional infall of surrounding matter. We hope that the 
{\em evolved} clusters will get at least closer to the {\em ideal} 
configuration of perfect equilibrium with no perturbing substructure.
The fact that we are missing part of the infalling matter should not 
affect the profiles at very small scales,but might cause some steepening 
in the outer parts of the density profiles.

\begin{figure}
\epsfxsize=\hsize\epsffile{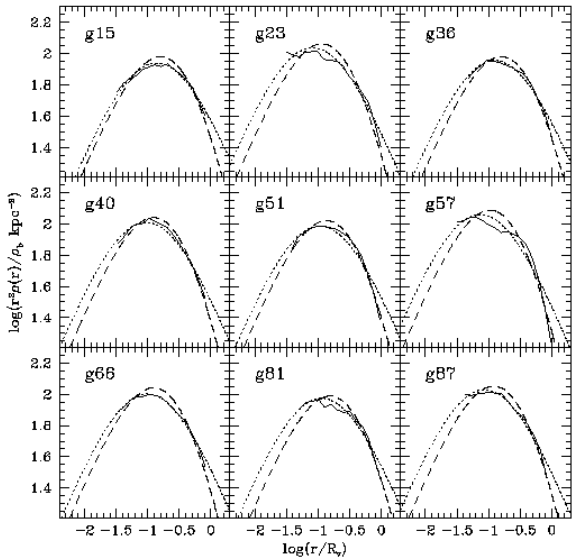}
\caption{The evolved (EVO) density profiles of our simulations are
plotted with their best fitting analytical NFW and HER models. Lines
are as in Figure~\ref{fig:adenfit}. Only the fitted range of each profile
(that is for radii larger than the radius enclosing 500 particles) is shown.}
\label{fig:edenfit}
\end{figure}

\begin{figure}
\epsfxsize=\hsize\epsffile{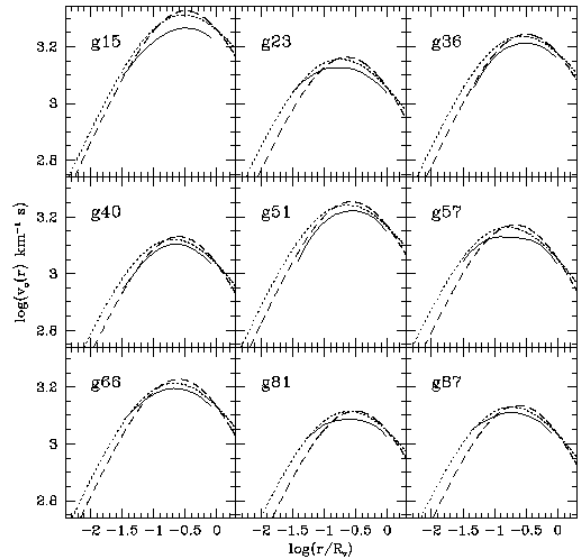}
\caption{Circular velocity profiles for the EVO outputs together with
their best fitting analitical NWF and HER profiles. Only the fitted 
part of each profile is shown.}
\label{fig:evcfit}
\end{figure}

At $t = 2 t_H$ we repeated the fitting analysis performed on the
original clusters. 
In order to obtain a {\em mean} evolved profile
we averaged the last four outputs, corresponding to times from 
$t=1.875 t_H$ to $t=2t_H$. We will name these profiles EVO (for 
evolved). On these we tried again both the NFW and HER fits.
We noticed that the density profiles of most halos tend to
flatten at very small radii as we evolve them beyond $t_H$.
It is likely that this effect is due to collisional relaxation,
or to the inaccuracy of the force evaluation of the code.
We therefore decided not to use the very inner part of the profiles 
in the fitting procedure. Specifically, the fits have been made only 
at radii larger than the radius enclosing 500 particles.
The density and circular velocity profiles of the evolved halos,
together with the best fits for the NFW and HER models are
presented in Figures~\ref{fig:edenfit} and \ref{fig:evcfit}. 

\begin{figure}
\epsfxsize=\hsize\epsffile{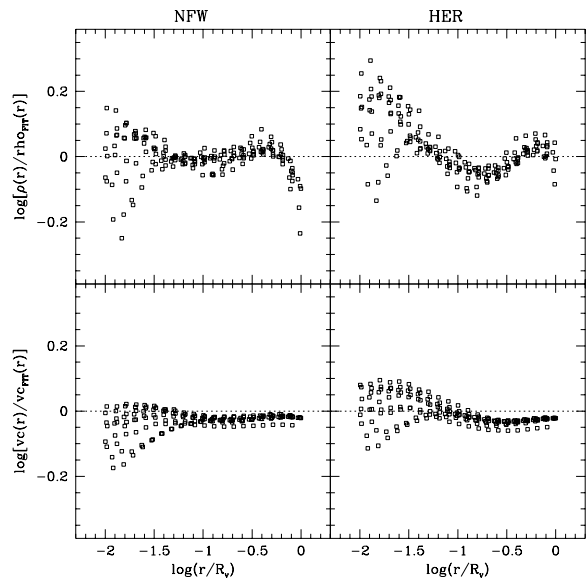}
\caption{Residuals of the fits to the EVO profiles in the previous two figure.}
\label{fig:evores}
\end{figure}

Halos evolved until $t=2t_H$ look much more dynamically relaxed than 
the original ones: very little substructure has survived within 
the virial radius. Moreover, the power law density profiles exhibited 
at intermediate radii by some halos in the previous configurations 
have disappeared, and all halos have now smoothly curved profiles. 
The residuals of the fits are shown in Figure~\ref{fig:evores}.
The scatter in the residuals is lower than before for the density, 
and about equal for the circular velocity; considering all the points 
down to the softening radius, the $1\sigma$ dispersion is 14\% (NFW) 
and 20\% (HER) for the density, and 7\% (NFW) and 10\% (HER) for the 
circular velocity. 
The improvement in the density fit shows that part of the scatter was 
due to the presence of substructure; as a halo approaches dynamical 
equilibrium its density is more accurately described by the NFW and 
HER profiles.
We conclude from this test that the NFW and HER models are indeed 
good fitting formulae for the equilibrium configuration of the density
and circular velocity of our dark matter halos.

\subsection{Trend with mass}\label{sec:mtrend}

In their N-body simulations of CDM dark matter halos, NFW recently 
found that more massive halos appear less centrally 
concentrated than less massive ones. NFW interpret this trend 
in terms of a dependence on the mean density of the universe 
at the time of halo formation, defined as in Lacey \& Cole (1993).
The trend they find is weak, and requires a large range of halo 
masses in order to be properly detected.
CL found a similar trend for the scale radius 
of their halos, formed in scale-free Einstein-de Sitter universes 
of the kind we are discussing.

\begin{figure}
\epsfxsize=\hsize\epsffile{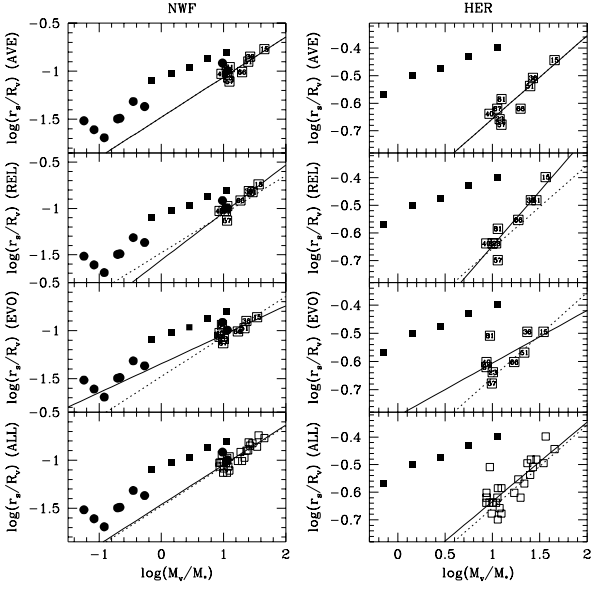}
\caption{Trend of the scale radius $x_s$ versus the normalized mass
of the halos, for different dynamical configurations. Left panels
refer to the NFW profile, right panels to the HER profile. Note the
wider mass range in the left panels. The meaning of different symbols 
and curves is given in the text.}
\label{fig:mtrend}
\end{figure}

In Figure~\ref{fig:mtrend} we show the result of the same test on 
our halos: the normalized scale radius $x_s$ is plotted versus the 
halo mass. Masses are normalized by the characteristic mass $M_*$, 
which is naturally defined at every time as the linear mass on the 
scale currently reaching the non-linear regime:
\be
M_*(t) = {4 \over 3} \pi R_*^3 \rho_b(t).
\ee
In the equation $\rho_b$ is the mean density of the universe, 
and $R_*$ is such that the linear density contrast on scale $R_*$ is 
$\delta(R_*) = 1.69$, the linear extrapolation to the collapse of a 
spherical top-hat. Recalling that the mass variance goes as 
$\delta_M \propto R^{-(3 + n)}$ for a scale free power spectrum
$P(k) \propto k^n$, and with our current normalization 
$\delta_M(8 h^{-1}$~Mpc$) = 0.63$, the value of $M_*$ for our
simulations is $M_* = 6.16\times 10^{13}$ M$_{\sun}$ 
at the present time (for $h=0.5$).

Open squares refer to our results; each number labels a halo. 
The first three rows show points corresponding to the AVE, REL 
and EVO configuration; the last row plots all points together.
The black circles refer to simulations of a $n=-1$ scale free 
Einstein-De Sitter universe (Navarro 1996), while the black 
squares are the results of CL. Note the wider mass range
covered by the Navarro (1996) results. Our data show that indeed 
more massive halos have a larger scale radius, but since they
cover a limited mass range we cannot make a very definite statement
on the slope of the relation. The solid lines are the best fit to
the points in each panel. The dotted line is the best fit to the 
AVE points, repeated in all panels to show how this is also
a reasonable slope for all configuration. Its value is $0.4$ and
$0.3$ for the NFW and HER points respectively.

For a given halo mass there are significant variations in the value 
of $x_s$ both within the same configuration and between different 
configurations. For the combined data the $1\sigma$ dispersion in 
scale radius is $14\%$ (NFW) and $11\%$ (HER). This means that halos 
best fit by the same model profile can differ in mass by a factor of 
two at a $2\sigma$ level.
Given this scatter, the points from Navarro are in good agreement
with ours in the range tested by our simulations. The points from CL 
instead are systematically higher: their halos seem less centrally 
concentrated than ours, and our relation is slightly steeper than theirs. 
Such a discrepancy could be due to the lower force and mass resolution
of the simulations used by CL: as a consequence, they did not push the
fit to radii as small as ours. To test this possible explanation, we 
repeated the fit on our {\em average} profiles two more times, using
for the fit the narrower ranges 
$\log(r/R_v) \in [-1.5,0]$ and $\log(r/R_v) \in [-1,0]$; these choices
bracket that of CL. We found that limiting the fit in this way does
indeed cause generally higher estimates for the scale radius, so we 
could marginally match the points of CL. We also found that fitting 
a narrower portion of the profiles causes a bigger scatter in the 
relation, to the point that, in the case $\log(r/R_v) \in [-1,0]$, 
most of the correlation between $r_s$ and mass was lost in 
our sample. Our understanding of this effect is the following. 
In limiting the fit to the outer part of the profile one becomes 
more and more sensitive to density enhancements at $r \approx R_v$, 
caused by merging substructure. These will both add noise and flatten 
the density profiles and thus will bias the fit towards higher values 
of scale radius and worsen the correlation with the halo mass.

A concurrent possibility for the discrepancy between us and CL 
could be that our halos have, perhaps, different average dynamical 
configurations than halos in the CL sample. 
To test this we repeated the fit using the most perturbed configuration 
of each halo: we did find a somewhat bigger scatter in the value of 
$r_s$ at low masses, but the overall trend came out similar to 
that found for the AVE and REL configurations. 
Further reasons could be the different criteria used to select the
halo centres and the best fitting value of $x_s$, or the fact that 
the number of timesteps used to evolve the simulations of CL may be 
too small to properly integrate orbits near the cluster centre.
Finally, we also tried to first average the density profiles 
$\rho(r/R_v)$ of halos of similar mass, and then fit the result 
with the analytical models, as CL do, but found no difference 
in our results.

To test the consistency of our results, we may use the fact that
our halos come from a self-similar universe. Therefore, we should 
be able to reproduce the relation between scale radius and halo 
mass using halos picked up from less evolved outputs.
We limited our choice to all outputs with $z<0.5$, to retain sufficient
mass and force resolution (we recall that our gravitational softening 
is fixed in proper coordinates, so that e.g. at $z=0.5$ the profiles 
are resolved down to $\log(r_{min}/R_v) \simeq -1.5$).
We found that, on average, less evolved halos require larger values for 
the scale radius; however, the difference between these and the 
results shown in Figure~\ref{fig:mtrend} is consistent with the effect
expected by fitting on a narrower range of radii, as explained above.

From these tests we conclude that indeed more massive halos have
on average flatter density profiles than less massive ones.
Although our mass range is only roughly a factor of six, the trend 
is statistically very significant: a Spearman rank correlation test 
shows in fact that the probability of a chance correlation for our 
points is 0.4\%, 0.2\%, 4.3\% for the AVE, REL and EVO data, and 
less than 0.1\% for the combined data. 
The relation suffers from a moderately large scatter due both to 
changes in the halo dynamics and to fluctuations in the halo
population. This fact, and the scatter between the present and 
previous results, make a precise determination of this relation 
difficult. For these reasons, we think that further study is 
needed before using it for applications.

\section{Other topics}

\subsection {Dynamical Equilibrium and mass estimation}\label{sec:dyneq}

We will now discuss the issue of equilibrium put forward in
Section~\ref{sec:evol}. Do our halos reach dynamical 
equilibrium? And how does merging affect this equilibrium
and the mass one estimates using the equilibrium hypothesis?

To answer this we use the Jeans' equation for the dynamical 
equilibrium of a collisionless system. For a static spherical 
system, it takes the form:
\ba
\nonumber
{d \over dr}\left[\rho(r) \sigma^2_r(r)\right] &+&
{\rho(r) \over r}\left[2\sigma^2_r(r) - \sigma^2_t(r)\right] \\
&=& -{G \rho(r) M(r)\over r^2 (1 + \epsilon^2/r^2)^{3/2}}
\label{eq:jeans}
\ea
where $G$ is Newton's constant, $\rho$ is the density at distance
$r$ from the cluster centre, $\sigma_r$ is the radial velocity 
dispersion and $\sigma_t$ is the tangential velocity dispersion
defined above. The extra factor $(1 + \epsilon^2/r^2)^{3/2}$
takes into account the fact that the Newtonian force is softened
at small radii; to first order we approximated the spline softening
with a Plummer softening $\epsilon = s/1.4$.
Solving for the mass gives 
\ba
\nonumber
M_E(r) = &-&{r \sigma_r^2(r) (1 + \epsilon^2/r^2)^{3/2} \over G} \\
&\times& \left[{d\ln\rho \over d\ln r} + {d\ln \sigma_r^2 \over d\ln r} 
+ 2\beta(r)\right],
\label{eq:mest}
\ea
where $\beta(r)$ is the anisotropy parameter defined above. 
We called the mass $M_E(r)$ because it is an estimate of the 
true mass within $r$.
We can compute the {\em rhs} of Equation~(\ref{eq:mest}) numerically, 
and compare $M_E(r)$ with the actual mass at different radii. 
Moreover, we can make different assumptions or approximations
by dropping one or more terms. This measures the sensitivity of 
the result to each term and quantifies the error made in estimating 
the true cluster mass. 

The simplest form of Equation~(\ref{eq:mest}) is obtained if we 
approximate the halo as a Singular Isothermal Sphere (SIS):
\be
M_{E1}(r) = {2 r v_{rms}^2(r) (1 + \epsilon^2/r^2)^{3/2} \over G}.
\ee
We allow the one dimensional {\em rms} velocity $v_{rms}^2 \equiv \sigma_r^2$ 
to vary with $r$ as in Figure~\ref{fig:vrmsdata}, and so use a 
different SIS at each radius. This expression in some sense is nothing 
more than the Virial Theorem, in that it relates the potential energy 
of the system to its kinetic energy. 

As a second estimate we assumed a non-singular Isothermal Sphere, 
and used the {\em actual} slope of the density profile:
\be
M_{E2}(r) = -{r v_{rms}^2(r) (1 + \epsilon^2/r^2)^{3/2} \over G} 
{d\ln\rho \over d\ln r}.
\ee

We can go further, and drop the isothermal approximation, still 
assuming that the velocity field is isotropic. In such a case 
Equation~(\ref{eq:mest}) reduces to
\be
M_{E3}(r) = -{r \sigma_r^2(r) (1 + \epsilon^2/r^2)^{3/2} \over G}
\left[{d\ln\rho \over d\ln r} + {d\ln \sigma_r^2 \over d\ln r}\right];
\ee
finally, we used the complete expression Equation~(\ref{eq:mest}), 
that is we assumed a general static and non rotating spherical 
system, with anisotropic velocity dispersion, and call it $M_{E4}$.
All these estimates rely on the assumption of spherical symmetry, 
which is not a very good approximation in at least half of 
our halos. Therefore, the results that we obtain here give only 
a rough idea of the role that equilibrium itself, and each term of 
Equation~(\ref{eq:mest}), play in the estimation of the halo mass.

\begin{figure*}
\epsfxsize=\hsize\epsffile{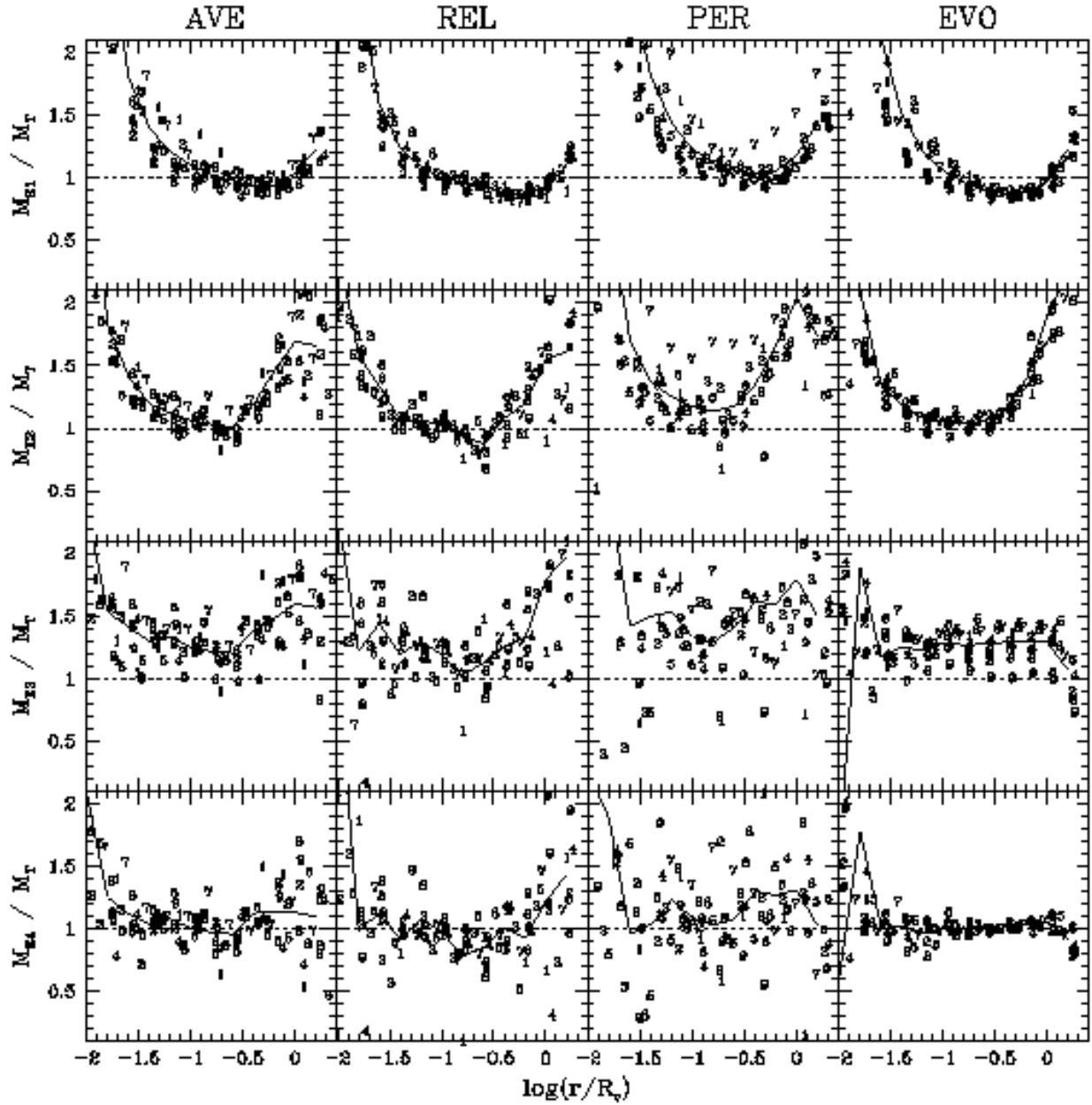}
\caption{Test of the halo dynamical equilibrium. The ratio of the mass
estimated by the Jeans' Equation $M_E$ to the true mass $M_T$ is
plotted versus the radial coordinate. Each line refers to a different 
estimator (explained in the text), and each column to a different 
dynamical configuration. Data for all halos are plotted in each panel,
labelled by number. The solid lines show the distribution mean values.}
\label{fig:mest}
\end{figure*}

We first verified the Jeans' equation on the {\em average} profiles, 
shown in Section~\ref{sec:profdata}. The idea is that different times 
have roughly the same dynamical importance, and that small variations 
between profiles are mostly due to transients in the cluster evolution.
To keep to a minimum the noise introduced by the numerical evaluation 
of the two derivatives, we smoothed the estimations with a Gaussian 
filter of logarithmic width 0.1.
The results of the four estimates $M_E$ are shown on the first column
of Figure~\ref{fig:mest}; each number labels a halo.

The main points to note are:
\begin{enumerate}
\item 
The analytically most accurate estimate, $M_{E4}$, shows that the dark 
matter halos are in reasonably good dynamical equilibrium at least 
within roughly one third of the virial radius. At larger radii the 
estimate is still fairly good but becomes more noisy due to the spikes
produced on the density profiles by infalling matter. 
At the virial radius the ratio between the estimated and true mass 
is $M_{E4}/M_T = 1.1 \pm 0.38$ ($1\sigma$ of the distribution).
At very small radii the noise is just Poissonian.

\item
Neglecting the velocity anisotropy makes the mass estimate larger
than the true mass, especially at large radii where $\beta(r)$ is 
significantly non zero due to predominance of radial orbits.
The large scatter in $M_{E3}$ and $M_{E4}$ is partly due to the noise 
introduced by the numerical differentiation.

\item
Although the SIS model gives wrong results at most radii, the ratio
$M_{E1}/M_T$ follows a clear trend crossing one around the virial 
radius. The different approximations in this estimate seem 
to introduce errors which balance one another at the cluster boundary, 
where this form of the Virial Theorem turns out to be quite accurate.
The mean ratio at the virial radius is in fact
$M_{E1}/M_T = 1.04 \pm 0.08$.

\item
Using the actual density slope in the isothermal model does not improve
the simple SIS estimate. Although at small radii $M_{E2}$ moves in the
right direction, at large radii it does not, and in general the true 
mass is overestimated everywhere but in a narrow region at intermediate 
radii.
\end{enumerate}

We now ask how these estimates $M_E$ would change if we applied them 
to different dynamical configurations of the clusters. For example,
we might expect a mass estimate made on a relaxed configuration to
be more accurate and less noisy than an average estimate, or an 
estimate made on a perturbed state to be higher than the true mass,
and also to have a larger scatter.
To test this we repeated the same analysis using for each cluster 
the most {\em relaxed} and the most {\em perturbed} outputs, defined 
in Section~\ref{sec:profdata}, and also using the {\em evolved} outputs
presented in Section~\ref{sec:evo}.
The result is shown in the second, third and fourth columns of
Figure~\ref{fig:mest}.

Here we can note the following things.
\begin{enumerate}
\item
All $M_E$ give similar results when applied to the averaged or to the 
most relaxed outputs. However, in the latter case they are slightly 
but systematically lower. They are also slightly noisier, probably 
because obtained from just one profile. The SIS model is still
good at the virial radius.

\item
All the estimators deal significantly worse when applied to the most 
perturbed outputs. Departures from dynamical equilibrium cause a 
general {\em overestimation} of the true mass, due to the higher 
velocity dispersion; the scatter in the measures is also larger.
However, $M_{E4}$ is still fairly consistent with the true mass, 
due to the large estimation noise. The SIS model fails in this 
case: $M_{E1}$ and $M_T$ differ by more than $1\sigma$.

\item
The halos evolved to $2t_H$ are indeed in very good dynamical
equilibrium: the estimate $M_{E4}$ is in excellent agreement with 
the true mass at all radii, with very small scatter. At the virial
radius the estimate is $M_{E4}/M_T = 1.05 \pm 0.05$. This tells us
that the assumption of a spherical system is actually a very good
approximation for this problem. The SIS estimate is again accurate 
at the virial radius, with $M_{E1}/M_T = 0.98 \pm 0.07$.

\item
Velocity dispersion anisotropies are important at all radii. 
Neglecting them leads to an average mass overestimation of 30\%.
\end{enumerate}

Of course in real clusters many facts make a comparison like this
more difficult. We used three dimensional profiles instead of 
projected ones, and did not include any of the observational 
errors present in the data. Moreover, in true clusters one 
observes the galaxy distribution, which may well not be tracing the 
underlying dark matter potential. Nevertheless, we believe that the 
test we performed here is still very instructive, since it can give 
us some insight of how the dynamical properties of halos can change 
during its formation, and to what extent simplifying approximations 
on the halo structure may be justified.

We may now ask what is the relative occurence of different dynamical
configurations in the cluster formation process: how often are the
halos relaxed? Or perturbed? We can roughly estimate this by looking
at Figure~\ref{fig:evol}. We saw that the virial {\em rms} velocity of the 
halos reaches a peak during each merging event, and decreases to a 
minimum after it. Let us count as {\em perturbed} all the outputs 
which, during a major merging event, have an {\em rms} velocity closer
to the peak value than to the following minimum. Analogously, let us 
count as {\em relaxed} the other outputs.
From the figure we can roughly say that our dark matter halos spend 
64\% of their time being relaxed and the other 36\% of the time
being perturbed. This means that a mass estimate done with one of
the above estimators will be likely biased upward with a probability 
of one third. The values shown in the third column of the figure 
should be regarded as a very rough upper limit, since they were 
computed on the {\em most} perturbed outputs only. 

\subsection {Consequences on Cosmology: Gravitational Lensing}

An interesting application of the present study is the comparison 
of the mass and density profiles of the simulations with those 
coming from observations of weak and strong gravitational lensing 
in rich clusters of galaxies. In particular, since the profiles 
are resolved down to few tens of kpc from the centre, these simulations
can be compared to the mass estimates at very small radii provided by 
giant arcs.

To compare the observed mass estimates to the simulated halos, we 
considered the simulations at different times, and selected from 
each output the high resolution particles forming the halo and the 
central part of the simulation. These particles were projected 100 
times along random lines of sight (l.o.s.), and projected profiles 
were computed in equally spaced logarithmic bins, following the same 
procedure used for the radial profiles. The mean values and 
dispersions in each bin of the profiles were finally computed.
In this procedure we neglected the low resolution particles that 
provide the tidal field of the simulations. Since the high 
resolution particles usually provide most of the mass out to two 
to three times the halo virial radius, neglecting the background 
particles has a negligible effect on the projected quantities, as
we verified.

\begin{figure}
\epsfxsize=\hsize\epsffile{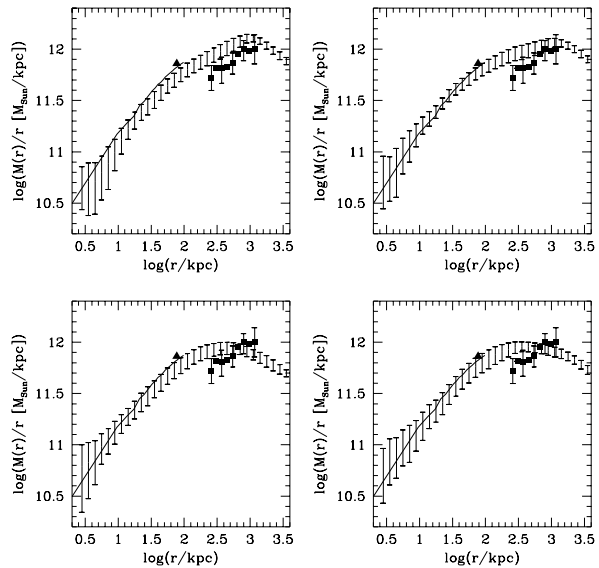}
\caption{mass profiles for A2218 plus four profiles from simulations.
The observed profile is repeated in each panel: the solid curve is a 
mass model inferred by combined arc and arclets observations 
(Kneib et al. 1995). The triangle is an estimate from a giant arc 
(Miralda Escud\`e \& Babul 1995). The solid squares come from a weak 
lensing model (Squires et al. 1995). The error bars indicate the
$\pm 1\sigma$ of the distribution of 100 random projections of
a simulated halo. The upper panels used two different outputs of
halo g15, the lower panels two outputs of halo g36.}
\label{fig:a2218}
\end{figure}

\begin{figure}
\epsfxsize=\hsize\epsffile{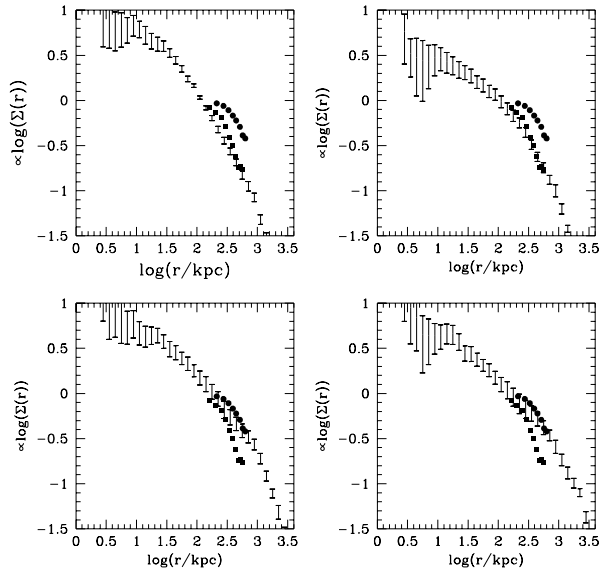}
\caption{Projected density profiles for 1455+22 (squares) and 0016+16 
(circles) in each panel, plus four different projected simulations. 
The data come from weak
lensing observations (Smail et al. 1995). In the two upper panels 
the simulated halos try to match the lower observed curve; viceversa
in the lower two panels. Normalization of the projected density is
arbitrary. Only the shape is to be considered. Simulation outputs
are taken from halo g57 and g51 in the upper panels, and from halo
g15 in the lower panels.}
\label{fig:smail}
\end{figure}

We did not constrain the simulated halos to have the same velocity
dispersion $\sigma_{los}$ as the observations for two reasons. 
First, the observed velocity dispersion is often computed as 
the best fit to some model (isothermal halo, $\beta$ model or 
others) and so is not measured in the same way as done for the 
simulated halos. Second, we want to leave $\sigma_{los}$ as a 
free parameter, in order to give a very crude estimate of the total
mass and velocity dispersion of the observed cluster based on the 
shape of their profiles.
We show in Figures~\ref{fig:a2218} and \ref{fig:smail} the
halos providing, by eye, the best fit to the observations.

Figure~\ref{fig:a2218} shows in each panel the mass profile of the rich 
cluster A2218, as inferred from different observations (details are
given in the Figure Caption).
The weak lensing measure is only a lower limit on the mass, and in
fact it falls slightly lower than the other estimates.
These observations are compared to four outputs of the simulations
(indicated by the thin error bars), two from each of the two most 
massive halos in the simulations
The error bars are centred on their mean projected mass, and 
enclose $\pm 1\sigma$ of the distribution of the 100 projections. 
The line of sight velocity dispersion measured within the Abell
radius ($R_{Ab} = 3$ Mpc), for the four outputs shown, are, 
from left to right and from top to bottom: 
$\sigma_{los}(R_{Ab}) = 1240$, $1230$, $950$, $950$ km s$^{-1}$. 
The observed estimate for A2218 is $\sigma_{los} = 1370$ 
km s$^{-1}$, but it refers to a smaller central region.

A better fit to the data is provided by the simulations in the upper 
panels, which come from a more massive halo and correspond to perturbed
dynamical configurations. 
The Abell mass (i.e. the mass within a
sphere of radius $R_{Ab}$) 
of these two outputs is equal to their virial mass and is of order 
of $M_{Ab} \simeq 2.2 \times 10^{15} M_{\sun}$. 
The {\em projected} mass within the Abell radius is slightly
bigger: $M = 2.5 \times 10^{15} M_{\sun}$.
The simulations in the lower panels fail instead to match the weak 
lensing lower limit on the mass at $\approx 1$ Mpc. 
The Abell mass (projected mass) for the halos in the lower panels is 
$M_{Ab} = 1.35 \times 10^{15} (1.8 \times 10^{15}) M_{\sun}$ (left) and 
$1.5 \times 10^{15} (2.1 \times 10^{15}) M_{\sun}$ (right). If we 
believe that the simulated halos are a reasonable representation
of a bright cluster like A2218, then their projected Abell mass 
can give an idea of the mass of A2218:
$2 \times 10^{15} M_{\sun} \leq M_{Ab}(A2218) 
\approx 2.5 \times 10^{15} M_{\sun}$.

In Figure~\ref{fig:smail} we plotted the projected density profiles
observed for the two distant clusters 1455+22 ($z = 0.26$, solid
squares) and 0016+16 ($z = 0.55$, solid circles).
The measures come from weak lensing observations, and the
normalization is not given. Therefore we only tried to fit the shape 
of the profile and chose the appropriate normalization in each panel. 
Even so, we can roughly guess the total mass and $\sigma_{los}$ of
these clusters by using the fact that halos of different mass have
profiles with different shape, as shown in Section~\ref{sec:mtrend}.
The error bars have the same meaning as in the previous figure, and
they now refer to outputs from three different halos. 

The velocity dispersion of the two upper halos is $\sigma_{los} = 
720$ km s$^{-1}$ (left) and $\sigma_{los} =1070$ km s$^{-1}$ (right). 
Their Abell mass (projected mass) is respectively $M_{Ab} = 8.0 
\times 10^{14} (8.4 \times 10^{14}) M_{\sun}$ and $M_{Ab} = 1.5 
\times 10^{15} (1.6 \times 10^{15}) M_{\sun}$. Although 
these values differ by almost a factor of two, their profiles fit 
equally well the observed profile of 1455+22, since the error bars 
on the observations are typically a factor of five. Their mass can 
perhaps bracket the mass of this cluster.
In the lower panels the simulations match instead cluster 0016+16.
The velocity dispersion of the halos are 
$\sigma_{los} =1230$ km s$^{-1}$ (left) and $\sigma_{los} =1150$ 
km s$^{-1}$ (right), similar to that estimated for the cluster.
Their Abell mass (projected mass) is respectively
$M_{Ab} = 2.2 \times 10^{15} (2.5 \times 10^{15}) M_{\sun}$ and 
$M_{Ab} = 2.4 \times 10^{15} (2.8 \times 10^{15}) M_{\sun}$. 
Again these can give an order of magnitude estimate of the total 
mass of the cluster.

\section{Summary and Conclusions}

In this paper we have used high resolution $N$-body simulations
to study the structure and dynamical evolution of the dark matter 
halos of galaxy clusters. 

Firstly we performed extensive numerical tests of some parameters 
that define the simulation setup and determine the resolution of 
the results. These are the number of particles forming the final 
halo, the gravitational softening parameter $s$ which sets the small 
scale cutoff of gravity, and the tolerance parameter that determines 
the accuracy of the time integration of the system. From these tests 
we could assess the accuracy limits of our simulations, and so choose 
parameter values appropriate to the dynamical resolution we required.
We then obtained a set of nine dark matter halos, resolved on average 
by $\approx 20000$ particles each, with an effective force resolution 
of $\simeq 25$ kpc.

We studied the formation process of these halos and discussed
their dynamical equilibrium and departures from this equilibrium
during their evolution. We analyzed the density and velocity field 
of the dark matter, especially in the central region of the halos, 
by using radial profiles of these quantities.
We tried two analytical fits to these profiles, and estimated their 
performance and range of applicability.
Under different approximations for the dynamics of the system, 
we tested the accuracy of the Jeans' equation in estimating the 
halo mass within different radii.
We finally compared the dark matter profiles of the simulated 
halos to those inferred from recent observations of gravitational
arcs, arclets and background distortions in rich clusters of 
galaxies.

Our main results are the following.
\begin{enumerate}
\item
The halo formation process can be simply schematized as an
alternation of merging phases and relaxation phases. Halos spend on
average one third of their evolution in perturbed configurations,
and the lasting two thirds in relaxed configurations. During merging
the halo increases its total mass by $20\%$ to $100\%$; its
velocity dispersion also increases by $\approx 15\%$ to $40\%$ due 
to the velocity of the infalling lumps. After the first passage of
a lump through the centre of the main halo, the latter 
starts to relax, the velocity dispersion decreases towards its
equilibrium value, and substructure is erased.
During these phases, halo densities can vary by up to a factor
of two, circular velocities by up to 60\% and radial velocity
dispersions by up to 70\%.

\item
The average configuration of a simulated halo is not isothermal.
If we call $\alpha$ the local logarithmic slope of the density
profile: $\rho(r) \propto r^{-\alpha}$, appropriate values for
the simulations are $\alpha \simeq -1$ at the smaller radii, 
and $\alpha \simeq -3$ to $-4$ around the virial radius.
The corresponding circular velocity profiles, 
$v_c(r) = (GM(r)/r)^{1/2}$, increase from the centre outwards,
reach a maximum value and start decreasing before the virial radius,
although only by a small amount, 10\% to 25\% of the peak value;
a similar trend is shown by the {\em rms} velocity.
The velocity dispersion within halos is anisotropic; orbits are more
elongated at larger radii, but almost isotropic at smaller radii.
\item
The analytic models proposed by NFW and HER fit the 
dynamically averaged halo profiles with good accuracy: the {\em rms} 
residuals for the density are 17\% and 26\% respectively. 
For the circular velocities they are 6\% and 10\%.
The two models provide an even better fit to the simulations if
one allows the halos to evolve until they erase most of their 
substructure. However, systematics in the residuals show that
the fits are slightly too flat at small radii and too steep
at larger radii.

\item
More massive halos have on average flatter density profiles than
less massive ones, as indicated by the trend of the scale radius
in the analytic models. The scatter in the relation is such that
the same analytical profile can be the best fit for halos with 
mass differing by a factor of two at a $2\sigma$ level.
\item
The assumption of a static spherical system is always a very good 
approximation for estimating the halo mass using the Jeans' Equation.
The true mass can be correctly recovered during most of the evolution,
and although the scatter in the estimate reaches a factor of two in 
dynamically perturbed halos, the average estimated mass always agrees
with the true value. In this sense, the halos are always in 
approximate dynamical equilibrium within their virial or Abell radius. 
A simple virial model provides a very good estimate of the Abell mass 
(but not of the mass at smaller radii) for halos not in a merging phase.
\item
Finally, we found that our simulations produce dark matter halos that 
can match quite well the dark matter distribution of galaxy clusters
A2218, 1455+22 and 0016+16, recovered from observations of 
gravitational lensing in these clusters, on scales from few kpc to 
$\approx 1$ Mpc.
\end{enumerate}

\section*{ACKNOWLEDGEMENTS}

It is a pleasure to thank Julio Navarro for very helpful suggestions, 
discussions and comments at different stages of this project.
Thanks also to Bhuvnesh Jain for useful discussions.
Julio Navarro is gratefully acknowledged for supplying a copy of 
his tree-SPH code, and Wendy Groom for providing the time integrator.
Financial support for G.T. was provided by an EC-HCM fellowship.

\end{document}